\documentclass[lettersize,journal]{IEEEtran}
\usepackage{amsmath}
\usepackage{amssymb}
\usepackage{amsfonts}
\usepackage{algorithmic}
\usepackage{algorithm}
\usepackage{array}
\usepackage{textcomp}
\usepackage{stfloats}
\usepackage{url}
\usepackage{verbatim}
\usepackage{graphicx}
\usepackage{subcaption}

\usepackage{cite}
\usepackage{xcolor}
\usepackage{multirow}
\usepackage{adjustbox}
\usepackage{booktabs}

\newcommand{\bh}[1]{\hat{#1}}

\begin{document}

\title{Iteration-Free Candidate Pruning and an Index-Domain VLSI Architecture for Sort-Free MIMO Box Decoding up to 1024-QAM}

\author{Shengchun Yang, Amit Sravan Bora,~\IEEEmembership{Member,~IEEE,} Emil Matus, and Gerhard Fettweis,~\IEEEmembership{Fellow,~IEEE}
\thanks{This work has been submitted to the IEEE for possible publication.
Copyright may be transferred without notice, after which this version may
no longer be accessible.}
\thanks{The authors would like to thank the Federal Ministry of Research, Technology, and Space (BMFTR) for its support as part of the research program Communication Systems ``Souverän. Digital. Vernetzt.''. Joint project 6G-life, project identification number: 16KIS2413K.}
\thanks{Shengchun Yang, Emil Matus, and Gerhard Fettweis are with the Vodafone Chair Mobile Communications Systems, Technische Universität Dresden, 01062 Dresden, Germany. Amit Sravan Bora is with the G. S. Sanyal School of Telecommunications, Indian Institute of Technology Kharagpur, Kharagpur 721302, India. (e-mail: \{shengchun.yang, emil.matus, gerhard.fettweis\}@tu-dresden.de; amitbora@gssst.iitkgp.ac.in).}
}

\markboth{IEEE Transactions on Circuits and Systems--I: Regular Papers}%
{Yang \MakeLowercase{\textit{et al.}}: Iteration-Free Pruning and Index-Domain VLSI Architecture for Sort-Free MIMO Box Decoding}


\maketitle

\begin{abstract}
Wi-Fi~7 and 5G NR use constellations up to $4096$-QAM and $1024$-QAM at multi-gigabit peak rates. The MIMO detector must sustain these rates at a low cost. Linear MMSE detection meets these rates at low cost but loses several dB in SNR. $K$-Best detection recovers this loss, but its per-layer sorting grows with the modulation order and needs large sorting networks or deep pipelines at these rates. Box Decoding is a sort-free tree-search detector whose complexity does not depend on the QAM order, and its deterministic pruning keeps the cost fixed as the MIMO size grows. Single-step candidate pruning (SCP) is fully parallel but less accurate, while iterative candidate pruning (ICP) is more accurate, but its selection loop costs either throughput or area. This paper proposes Max-Min-Exchange Candidate Pruning (MMECP), which replaces the selection loop of ICP with a single max-min exchange without a feedback path and keeps \mbox{near-$K$-Best} performance. The paper also presents a VLSI architecture that represents each candidate by its integer constellation index, so that candidates and their distances are obtained by bit selection instead of operations on symbol values. A $4\times4$ detector synthesized in TSMC $16$\,nm detects one vector per cycle at $1$\,GHz for $64$- to $1024$-QAM and produces up to $40$\,Gb/s, while its gate count grows by $63\%$ over this range. Its normalized hardware efficiency is $2.3\times$, $12\times$ and $14\times$ that of the best compared hard-output detector at $64$-, $256$- and $1024$-QAM.
\end{abstract}


\section{Introduction}
Multiple-input multiple-output (MIMO) technology improves the spectral efficiency, link reliability, and capacity of modern wireless systems through spatial multiplexing and diversity~\cite{Rusek_MIMO_2013}. As standards advance from 5G toward 6G systems, the QAM order keeps increasing. Wi-Fi~7 (IEEE~802.11be) supports up to $4096$-QAM~\cite{ieee_80211be}, and 5G New Radio (NR) supports up to $1024$-QAM in the downlink since Release~17~\cite{3gpp_38214}. At peak rates of tens of Gb/s~\cite{ieee_80211be, itu_m2410}, the MIMO detector must sustain multi-gigabit throughput within the fixed processing time of these standards~\cite{3gpp_38214}.

MIMO detectors fall into three classes~\cite{MIMO_detection_Hanzo, MIMO_Larsson}. Maximum-Likelihood (ML) and Maximum A Posteriori (MAP) detection are optimal, but their complexity grows exponentially with the MIMO dimension and the modulation order. Linear detectors such as Zero-Forcing (ZF) and Minimum Mean-Square Error (MMSE) have low complexity but lose several dB in SNR compared with tree-search detection, especially in ill-conditioned channels. Tree-search detectors lie between these two classes. They operate on the QR-decomposed channel matrix, avoid explicit matrix inversion, and approach ML performance by exploring only a subset of the candidate paths.

Many tree-search detectors have been proposed~\cite{Bello_VSLI_Survey_2016, qrmmld_2004, cha_perm2025, New_SD_2025}. Sphere Decoding (SD)~\cite{sphere_viterbo} approaches ML performance, but its data-dependent complexity gives variable throughput and latency, which conflicts with the deterministic processing time required above. The LDPC codes of 5G NR and Wi-Fi recover much of the loss of sub-optimal detection, so a fixed complexity and a parallel architecture are preferred over near-ML accuracy~\cite{soft_fcsd_2025}. Reduced-complexity ML detection, $K$-Best and fixed-complexity SD are therefore used in multi-user MIMO (MU-MIMO) receivers \cite{mu_kbest_2020, mu_fsd_2022}, and reduced-complexity ML receivers are investigated for 3GPP 5G-Advanced systems \cite{mu_rml_2025}.

$K$-Best detection has been widely implemented in hardware \cite{K-best_JSAC, DKB_v1, KBest_VLSI, soft_output_2010, DKB_conf, k_best_ISCAS, k_best_lattice_reduction}, including radius-adaptive, reduced-complexity, and high-throughput architectures \cite{radius_adaptive_2010, multigigabit_2014, partial_noise}. By retaining the best $K$ candidates at each tree level, it achieves deterministic complexity, throughput, and latency. The fixed-complexity sphere decoder (FSD) is a related scheme that expands a fixed number of candidates per layer, but its complexity still grows with the QAM order \cite{soft_fcsd_2025, mu_fsd_2022}. However, retaining the best $K$ candidates requires sorting the expanded candidates at every layer, and the compare-and-swap count of a sorting network grows with $n\log^2 n$ and its depth with $\log^2 n$ for $n$ candidates~\cite{batcher_1968}. Hardware designs therefore fold the sort over several cycles, which divides the throughput~\cite{K-best_JSAC}, or replace it with an iterative bisection search over the candidate PEDs, which increases the area cost~\cite{sort_free_2024}. In both cases, the selection dominates the area and latency of a layer and limits the throughput per unit area. Its cost also grows with the modulation order \cite{sort_free_2024, k_best_sort_free}, which has led to dedicated sorting architectures \cite{KBest_Sorter_2017, Sorter_2023, multiway, merge_max_set}.

Sort-free detection is therefore attractive for high-order QAM. Selective Spanning with Fast Enumeration (SSFE)~\cite{ssfe_icc2008} expands a fixed number of nearest constellation points at each node without sorting, but its candidate count is controlled by the per-node spanning factors. This means that a large spanning factor increases the number of candidates per layer. To constrain the number of candidates and keep the sorting out of the detection layer, the distributed $K$-Best (DKB) detector~\cite{DKB_conf} gives a solution, which expands the candidate nodes on demand by Schnorr-Euchner enumeration with decision-boundary quantization~\cite{DKB_v1}, and selects the $K$ survivors in an iterative loop. Node-perturbation detectors~\cite{lund_np_2013} also avoid sorting, but their candidate expansion grows with the constellation.

Box Decoding~\cite{box} is a recent sort-free detector. At each layer, it expands a fixed box of $B=4$ nearest points around the ZF estimate of every surviving path, so its complexity is independent of the QAM order. On a vector DSP for 5G NR RedCap with $2\times2$ MIMO and $64$-QAM, it reaches near-$K$-Best performance with about $94\%$ fewer visited nodes per layer than a fully expanding $K$-Best detector~\cite{box}. Without pruning, however, the number of visited nodes grows exponentially with the MIMO size. $K$-Box~\cite{KBox} reaches $K$-Best performance by selecting the $K$ best of the $K\cdot B$ candidates per layer, but reintroduces a sorter. Probabilistic tree pruning~\cite{box_ptp} removes unlikely nodes with a statistically derived threshold and needs no sorting, but the number of surviving nodes depends on the SNR, so the hardware must be designed to suit the worst case. Deterministic candidate pruning~\cite{box-scp} keeps a fixed number of paths without sorting. Its single-step form (SCP) decides inside each Box cluster and is fully parallel, but loses accuracy against $K$-Best. Iterative Candidate Pruning (ICP) reaches $K$-Best performance with a selection loop similar to the on-demand selection of DKB. Its $K$ iteration steps depend on each other, so the loop either takes $K$ cycles per vector or, when unrolled, adds $K$ sequential selection stages with extra area and latency. DKB, \mbox{$K$-Box} and ICP therefore all reach the detection performance of $K$-Best and differ in hardware cost. \mbox{$K$-Box} needs a sorter, whereas DKB and ICP need $K$ dependent selection rounds, whose area and throughput are compared in Section~\ref{sec:ces_cmp}.

This work targets a fully pipelined, feed-forward detector that accepts one received vector per clock cycle with a fixed latency, so that a single instance reaches the multi-gigabit rates above at high-order QAM. The pruning must still keep near-$K$-Best accuracy at a fixed $K$.

To meet this target, this paper proposes an iteration-free pruning scheme and a VLSI architecture for sort-free Box Decoding with fixed throughput. The contributions are as follows.
\begin{itemize}
\item \emph{Iteration-free candidate pruning.} A Max-Min-Exchange Candidate Pruning (MMECP) scheme keeps $K$ paths per layer with a single max-min exchange, using $2K-1$ comparators, no sorter and no feedback path. It reaches the detection performance of Box-SICP, about $0.2$\,dB from $K$-Best at $64$- to $1024$-QAM.
\item \emph{Index-domain VLSI datapath with MUX-based pruning.} The detector represents each constellation coordinate by its integer index and keeps the datapath in this index QAM map $\tilde{\mathcal{A}}$ from the ZF estimate of each surviving path to the accumulation of the partial Euclidean distance (PED), computed with the $\ell_1$-norm. In $\tilde{\mathcal{A}}$, candidate generation is implemented with bit selection instead of additions and subtractions on symbol values, and the selection of SCP and MMECP reduces to a minimized Boolean function of six and eight bits that drives a multiplexer (MUX) network in place of a compare-and-swap network. The index width grows only by one bit per QAM step, and the selection logic is the same for every modulation order.
\item \emph{Implementation and evaluation.} A fixed-point $4\times4$ detector synthesized in $16$\,nm CMOS detects one vector per cycle at $1$\,GHz for $64$- to $1024$-QAM, delivering up to $40$\,Gb/s with $63\%$ gate-count growth over this range. The candidate expansion and selection blocks are compared in area and power between $\mathcal{A}$ and $\tilde{\mathcal{A}}$ and across SCP, MMECP, ICP and DKB. The full detector reaches $2.3\times$, $12\times$ and $14\times$ the normalized hardware efficiency of the best compared hard-output detector at $64$-, $256$- and $1024$-QAM, and $1.2\times$, $3.0\times$ and $3.5\times$ its $K$-normalized FOM.
\end{itemize}

The remainder of this paper is organized as follows. Section~\ref{sec:system_model} reviews the tree-based MIMO detection model, Box Decoding, and the deterministic pruning rules of SCP and ICP. Section~\ref{sec:pruning} presents the survival analysis, the proposed MMECP, and its sorting accuracy. Section~\ref{sec:Sim} reports the pruning complexity, the bit error rate (BER) performance, and the fixed-point quantization. Section~\ref{sec:idx_alg} reformulates the detection in the index QAM map, Section~\ref{sec:arch} details the proposed VLSI architecture, and Section~\ref{sec:hw_results} presents the synthesis-based comparison of the datapaths and pruning schemes and the comparison with the state of the art. Table~\ref{tab:notation} summarizes the notation used throughout.

\begin{table*}[t]
\caption{Notation and symbols.}
\label{tab:notation}
\centering
\begin{subtable}[t]{0.49\textwidth}
\vspace{0pt}
\centering
\footnotesize
\begin{tabular}{@{}ll@{}}
\toprule
\textbf{Notation} & \textbf{Description}\\
\midrule
\multicolumn{2}{@{}l}{\textit{Scalars and indices}}\\
$N$ & MIMO size ($N_{tx}=N_{rx}=N$)\\
$\mathcal A$,\ $\tilde{\mathcal A}$ & QAM map, index QAM map~(\ref{eq:idx_map})\\
$\Delta q,\ M$ & QAM spacing ($2$), max.\ index $\sqrt{|\mathcal A|}-1$\\
$\gamma$ & lowest QAM grid value $-M\Delta q/2$\\
$n,\ p$ & QAM grid indices (real, imag.)\\
$u,\ v$ & Box offsets (real, imag.)\\
$B,\ K$ & Box size ($4$), no.\ of survivors (Box clusters)\\
$i,\ k$ & layer index, Box cluster index\\
\midrule
\multicolumn{2}{@{}l}{\textit{Operators}}\\
$\mathrm{Box}(\cdot)$ & Box operation~(\ref{eq:box_def})\\
$\mathrm{PED}(\cdot)$ & PED calc.~(\ref{eqn:PEDa})\\
$\mathrm{Rule1}(\cdot),\ \mathrm{Rule2}(\cdot)$ & pruning Rule~1~(\ref{eqn: Rule1}), Rule~2~(\ref{eqn: Rule2})\\
\midrule
\multicolumn{2}{@{}l}{\textit{Index domain ($\tilde{\mathcal A}$, Sec.~\ref{sec:idx_alg})}}\\
$\rho_{i,j}$ & interference coeff.\ $r_{i,j}/(r_{i,i}\Delta q)$~(\ref{eq:idx_a})\\
$\lambda$ & offset $M(1+\mathrm{j})/\Delta q$~(\ref{eq:idx_a})\\
$\mathfrak i,\ \mathfrak f$ & integer, fractional parts of $\tilde a$\\
$\mathfrak p_n,\ \mathfrak p_o$ & negative, overflow position flags\\
$\omega^{\mathrm{re}},\ \omega^{\mathrm{im}}$ & overflow values $\mathfrak R\{\tilde a\}-M$, $\mathfrak I\{\tilde a\}-M$\\
\bottomrule
\end{tabular}
\caption{Parameters, indices, and operators}
\label{tab:notation_params}
\end{subtable}
\hfill
\begin{subtable}[t]{0.49\textwidth}
\vspace{0pt}
\centering
\footnotesize
\begin{tabular}{@{}ll@{}}
\toprule
\textbf{Notation} & \textbf{Description}\\
\midrule
\multicolumn{2}{@{}l}{\textit{Nodes and distances (Box cluster $k$, layer $i$)}}\\
$\hat{(\cdot)},\ \tilde{(\cdot)}$ & quantity in $\mathcal A$, quantity in $\tilde{\mathcal A}$\\
$\hat a_{k,i}$ & reference node~(\ref{eq:box_ai_alg})\\
$\hat x_{k,i}(l)$ & candidate node $\mathrm{Box}(\hat a_{k,i})$, $l=1,\dots,B$\\
$\hat\delta_{k,i}(l)$ & displacement $\hat a_{k,i}-\hat x_{k,i}(l)$\\
$\hat{\mathcal B}_{k,i}$ & Box cluster $\{\hat x_{k,i}(l)\}_{l=1}^B$\\
\midrule
\multicolumn{2}{@{}l}{\textit{Ranks (Sec.~\ref{sec:surv_analysis})}}\\
$\hat x^{(b)}_{k,i}$ & local rank $b$ in $\hat{\mathcal B}_{k,i}$; 1st/2nd winner for $b=1,2$\\
$\mathcal G^{(b)}_i$ & rank group $\{\hat x^{(b)}_{k,i}\}_{k=1}^K$\\
$(b,g)$ & rank pair: local rank $b$, group rank $g$\\
\midrule
\multicolumn{2}{@{}l}{\textit{Paths and metrics}}\\
$\hat X_{k,i}$ & partial path $\{\hat x_{k,j}\}_{j=i}^N$\\
$\hat d_{k,i}$ & path PED $\mathrm{PED}(\hat X_{k,i})$\\
$\hat{\mathbb X}_i$ & augmented path set $\{(\hat X_{k,i},\hat d_{k,i})\}_{k=1}^K$\\
$\hat W_{k,i}^{(b)}$ & winner path $(\hat X_{k,i+1},\hat x^{(b)}_{k,i})$, $b=1,2$\\
$\hat{\mathbb W}_i^{(b)}$ & augmented rank group $\mathcal G^{(b)}_i$, $b=1,2$\\
\bottomrule
\end{tabular}
\caption{Nodes, ranks, and paths}
\label{tab:notation_paths}
\end{subtable}
\end{table*}

\section{System Model and Box Decoding}
\label{sec:system_model}

\subsection{Tree-Based MIMO Detection}

Consider a spatially multiplexed $N \times N$ MIMO system operating over a Rayleigh flat-fading channel. Let $\mathbf{x} \in \mathbb{C}^N$ be the transmitted symbol vector drawn from a QAM constellation of size $|\mathcal{A}|$, and let $\mathbf{H} \in \mathbb{C}^{N \times N}$ denote the MIMO channel matrix. The received signal vector $\mathbf{y} \in \mathbb{C}^N$ can then be modeled as
\begin{equation}
\mathbf{y} = \mathbf{H}\mathbf{x} + \mathbf{w},
\end{equation}
where $\mathbf{w} \in \mathbb{C}^N$ is an additive white Gaussian noise (AWGN) vector with independent and identically distributed (i.i.d.) entries. Applying the QR decomposition $\mathbf{H} = \mathbf{Q}\mathbf{R}$, where $\mathbf{Q}$ is unitary and $\mathbf{R}$ is upper triangular, the system model is expressed as
\begin{equation}
\mathbf{y}' = \mathbf{Q}^H \mathbf{y} = \mathbf{R}\mathbf{x} + \mathbf{w}',
\label{equ:QR}
\end{equation}
where $\mathbf{w}' = \mathbf{Q}^H\mathbf{w}$ remains AWGN because $\mathbf{Q}$ is unitary. Since $\mathbf{R}$ is upper triangular, \eqref{equ:QR} can be detected layer by layer, which is the basis of tree-search detection. All detectors in this work operate on \eqref{equ:QR} and differ only in their candidate expansion and pruning.

\subsection{Box Decoding}
\begin{figure}[t]
        \centering
        \includegraphics{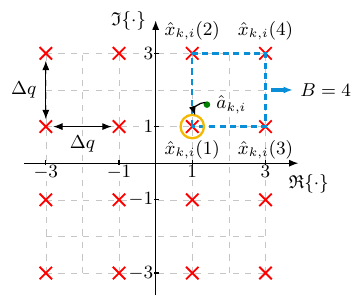}
        \caption{Box decoding example~\cite{box}.}
        \label{fig:Box_QAM_diagram}
        \vskip -0.5\baselineskip
\end{figure}
Box Decoding detects layer by layer from the top layer (root) $i = N$ down to the bottom layer (leaves) $i = 1$. At each layer, $K$ paths are selected to be retained for the next layer. They are indexed by $k\in\{1,\dots,K\}$, and the $k$-th partial path is $\bh{X}_{k,i+1} = \{\hat{x}_{k,j}\}_{j=i+1}^{N}$, where $\hat{x}_{k,j}$ is the estimated symbol at layer $j$ along the $k$-th parent path. At the top layer $i = N$, detection starts from an empty parent path, and the corresponding \textit{reference node} (also the ZF estimate) is given by $\hat a_{N} = y'_N / r_{N,N}$. At layer $i$, the reference node of the $k$-th parent path is obtained by canceling the interference of the detected symbols,
\begin{equation}
\hat {a}_{k,i}
=
\frac{1}{r_{i,i}}
\left(
y'_i
-
\sum_{j=i+1}^{N} r_{i,j}\hat{x}_{k,j}
\right).
\label{eq:box_ai_alg}
\end{equation}
Instead of expanding all constellation points of $\mathcal{A}$, Box Decoding generates only the $B$ points of a square region (``Box'') around $\hat a_{k,i}$, as shown in Fig.~\ref{fig:Box_QAM_diagram}.\par
The \textit{Box operation} $\mathrm{Box}(\cdot)$ quantizes the reference node (ZF estimate) to the lower-left corner of the box (highlighted in yellow in Fig. \ref{fig:Box_QAM_diagram}) and generates $B$ candidates around it on the QAM grid with spacing $\Delta q$, forming the \textit{Box cluster}
\begin{equation}
\hat{\mathcal B}_{k,i} = \mathrm{Box}(\hat a_{k,i}) = \{\hat{x}_{k,i}(l)\}_{l=1}^{B},
\label{eq:box_def}
\end{equation}
where each candidate is given as
\begin{equation}
\hat{x}_{k,i}(l) = \hat{x}^{\mathrm{re}}_{k,i}(u) + \mathrm{j}\,\hat{x}^{\mathrm{im}}_{k,i}(v),
\end{equation}
with
\begin{equation}
\begin{aligned}
\hat{x}^{\mathrm{re}}_{k,i}(u) &=
\left(\left\lfloor\frac{\mathfrak{R}\{\hat a_{k,i}\}-\gamma}{\Delta q}\right\rfloor + u \right)\Delta q+\gamma, \\
\hat{x}^{\mathrm{im}}_{k,i}(v) &=
\left(\left\lfloor\frac{\mathfrak{I}\{\hat a_{k,i}\}-\gamma}{\Delta q}\right\rfloor + v \right)\Delta q+\gamma,
\end{aligned}
\label{eqn:box_operation}
\end{equation}
where $l \in \{1,\dots,B\}$ indexes candidates in the Box cluster and $(u,v)$ are the real and imaginary grid offsets with $u,v \in \{0,\dots,\sqrt{B}-1\}$, ordered as $l=1+\sqrt{B}\,u+v$ so that $\hat x_{k,i}(1)$ is the lower-left corner of the box, called the \textit{starting candidate}.\par
The QAM grid takes the values $\gamma+n\,\Delta q$ on the real axis and $\gamma+p\,\Delta q$ on the imaginary axis, with QAM grid indices $n,p\in\{0,1,\dots,M\}$, where $M = \sqrt{|\mathcal{A}|} - 1$ is the maximum QAM grid index per dimension and $\gamma=-M\Delta q/2$ is the lowest value of the QAM grid on each axis. Subtracting $\gamma$ shifts the origin-centered grid to the first quadrant, so the floor returns the grid index of the lower-left corner of the box, and adding $\gamma$ maps the candidate back to the original QAM map. The floored index is clipped to $\{0,\dots,M-1\}$, so the box always lies inside the constellation. For the integer grid with $\Delta q=2$ used in this work, $\gamma=-M$, so subtracting the minimum value $\gamma$ is the same as adding the maximum value $M$ of the QAM grid.\par
After pruning, each surviving candidate $\hat{x}_{k,i} \in \hat{\mathcal B}_{k,i}$ is appended to its parent path $\bh{X}_{k,i+1}$ to form the new partial path $\bh{X}_{k,i}$, with updated PED
\begin{equation}
\bh{d}_{k,i}=\bh{d}_{k,i+1}+\left|y'_i-\sum_{j=i+1}^{N} r_{i,j}\hat{x}_{k,j}-r_{i,i}\hat{x}_{k,i}\right|^2.
\label{eqn:PEDa}
\end{equation}
The $K$ surviving paths are then collected into the augmented path state set
\begin{equation}
    \bh{\mathbb{X}}_i = \{(\bh{X}_{k,i}, \bh{d}_{k,i})\}_{k=1}^{K},
\end{equation}
where each tuple $(\bh{X}_{k,i}, \bh{d}_{k,i})$ pairs a path with its accumulated PED. The index $k$ labels the $K$ survivors at layer $i$ and is reassigned after the pruning at each layer\footnote{The index $k$ in $\bh{\mathbb{X}}_i$ need not match the parent index in $\bh{\mathbb{X}}_{i+1}$, since several survivors can share one parent while another parent leaves none. The survivors are relabeled $k=1,\dots,K$ at each layer.}.\par
One detection layer consists of \emph{candidate expansion}, i.e., the Box operation that generates the Box cluster, \emph{candidate selection}, i.e., the pruning that keeps $K$ paths, and the \emph{PED calculation}~\eqref{eqn:PEDa} of the candidates. The detection repeats these three steps layer by layer down to $i=1$, where the decoder outputs the path $\bh{X}_{\min}$ in $\bh{\mathbb{X}}_{1}$ with the minimum accumulated PED.

\subsection{Deterministic Pruning Strategies: SCP and ICP}
\label{sec:scp_icp_recap}
This subsection reviews the deterministic pruning rules and the SCP, ICP, and SICP$_t$ algorithms of~\cite{box-scp}. Let $\hat\delta_{k,i}(l)=\hat a_{k,i}-\hat{x}_{k,i}(l)$ denote the displacement between the reference node and the $l$-th candidate of Box cluster $k$ at layer~$i$. Since all $B$ candidates of a Box cluster share the same parent-path PED, their PEDs are ordered by $|\hat\delta_{k,i}(l)|^2$. The candidate with the smallest displacement, termed the \emph{first winner} $\hat{x}^{(1)}_{k,i}$, is found by two real-valued comparisons on the displacement $\hat\delta_{k,i}(1)$ to the box corner, known as \emph{Rule~1} and denoted $\mathrm{Rule1}(\cdot)$:
\begin{equation}
    \Delta q - 2\,\mathfrak{R}\{\hat\delta_{k,i}(1)\} \;\gtrless\;0,\qquad
    \Delta q - 2\,\mathfrak{I}\{\hat\delta_{k,i}(1)\} \;\gtrless\;0.
    \label{eqn: Rule1}
\end{equation}

\textbf{Single-Step Candidate Pruning (SCP)} applies Rule~1 to each of the $K$ Box clusters independently and keeps one candidate per Box cluster, so it is fully parallel and requires no comparisons across Box clusters. However, it may discard candidates that a global sort would keep, which is known as the \emph{optimal child-node skip problem}. To address this, \emph{Rule~2}, denoted $\mathrm{Rule2}(\cdot)$, orders the remaining candidates within a Box cluster by two further comparisons,
\begin{equation}
\begin{aligned}
\mathfrak{R}\{\hat\delta_{k,i}(1)\} - \mathfrak{I}\{\hat\delta_{k,i}(1)\} \;&\gtrless\; 0,\\
\mathfrak{R}\{\hat\delta_{k,i}(1)\} + \mathfrak{I}\{\hat\delta_{k,i}(1)\} \;&\gtrless\; \Delta q.
\end{aligned}
\label{eqn: Rule2}
\end{equation}
Together with Rule~1, it orders all $B$ candidates of a Box cluster without computing any PED. \par
\textbf{Iterative Candidate Pruning (ICP)} treats each Box cluster as a sorted list. At each iteration, it selects the candidate with the smallest accumulated PED and replaces it by the next candidate of the same Box cluster, until $K$ paths survive. ICP keeps the same survivors as global sorting and matches the detection performance of \mbox{$K$-Box}~\cite{KBox}. Its selection is similar to the on-demand expansion of DKB~\cite{DKB_conf}, restricted to the four children of a box. The iterations depend on each other and cannot run in parallel. The hybrid SCP--ICP variant SICP$_t$ applies ICP only in the first $t$ pruning layers and SCP in the remaining layers. SICP$_1$, referred to as \emph{Box-SICP} in the rest of this paper, closely approaches the BER of ICP, but even a single iterative layer prevents fully parallel execution. The proposed MMECP keeps the detection performance of \emph{Box-SICP} without the iterative loop.

\section{Proposed Max-Min-Exchange Candidate Pruning}
\label{sec:pruning}

\begin{figure*}[t]
    \centering
    \begin{subfigure}[b]{0.32\textwidth}
        \centering
        \includegraphics[width=\linewidth]{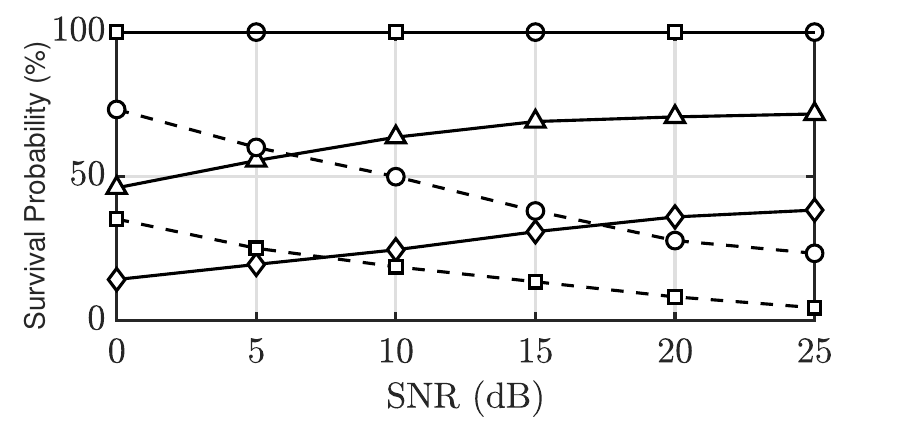}
        \caption{Survival probability}
        \label{fig:sur_prob}
    \end{subfigure}
    \hfill
    \begin{subfigure}[b]{0.32\textwidth}
        \centering
        \includegraphics[width=\linewidth]{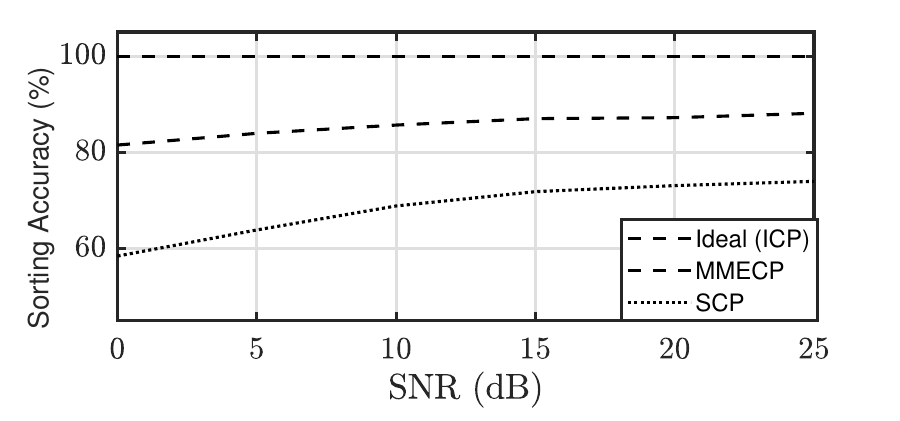}
        \caption{Layer-wise accuracy}
        \label{fig:ll_agr}
    \end{subfigure}
    \hfill
    \begin{subfigure}[b]{0.32\textwidth}
        \centering
        \includegraphics[width=\linewidth]{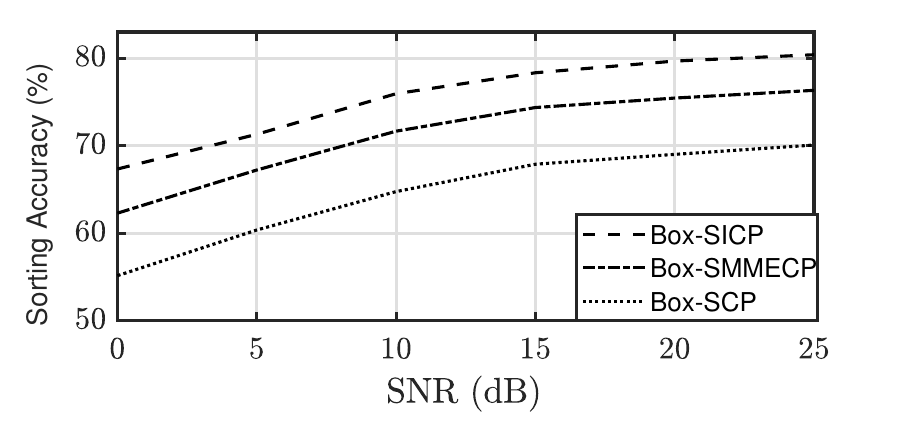}
        \caption{Tree-wise accuracy}
        \label{fig:tl_agr}
    \end{subfigure}
    \caption{Statistics for a $4\times4$ $64$-QAM detector: (a) survival probability of the candidates of group rank 1 $(\circ)$, 2 $(\square)$, 3 $(\triangle)$, and 4 $(\diamond)$ in the rank groups $\mathcal G^{(1)}$ (solid) and $\mathcal G^{(2)}$ (dashed) under global sorting; (b) layer-wise and (c) tree-wise sorting accuracy versus SNR.}
    \label{fig:agreement_curves}
\end{figure*}

\subsection{Survival Analysis of Ranked Candidates}
\label{sec:surv_analysis}
We first define two ranks. Candidates are ranked by PED in ascending order, so rank~1 denotes the smallest PED.
\begin{itemize}
    \item \emph{Local rank.} Rule~1 and Rule~2 order the candidates inside one Box cluster. Rule~1 gives the candidates of local rank~1 and local rank~4, and Rule~2 decides which of the two remaining candidates has local rank~2. The candidate of local rank~$b$ in Box cluster~$k$ is denoted $\hat x^{(b)}_{k,i}$, so $\hat x^{(1)}_{k,i}$ and $\hat x^{(2)}_{k,i}$ are the first and second winners.
    \item \emph{Group rank.} The candidates of local rank~$b$ across the $K$ Box clusters form the rank group $\mathcal G^{(b)}_i=\{\hat x^{(b)}_{k,i}\}_{k=1}^{K}$. Their parent paths differ, so they are ranked by their accumulated PEDs, from group rank~1 to group rank~$K$.
\end{itemize}
Each candidate is thus identified by its rank pair $(b,g)$ of local rank~$b$ and group rank~$g$. For example, $(b,1)$ is the candidate with the smallest accumulated PED in $\mathcal G^{(b)}_i$, and $(1,K)$ the one with the largest accumulated PED in $\mathcal G^{(1)}_i$. The SCP survivor set is $\mathcal G^{(1)}_i$. \emph{Global sorting} ranks all $KB$ candidates of a layer by their accumulated PEDs and keeps the $K$ smallest, as ICP does. A candidate \emph{survives} if global sorting keeps it.

Fig.~\ref{fig:sur_prob} shows the survival probability of the candidates in $\mathcal G^{(1)}_i$ and $\mathcal G^{(2)}_i$, obtained by Monte Carlo simulation of a $4\times4$ $64$-QAM system over an i.i.d.\ Rayleigh flat-fading channel. Two observations motivate MMECP:
\begin{enumerate}
    \item The rank groups $\mathcal G^{(3)}_i$ and $\mathcal G^{(4)}_i$, and the candidates $(2,3)$ and $(2,4)$ (not shown), have negligible survival probability across the SNR range and can be ignored during pruning.
    \item In $\mathcal G^{(1)}_i$, the candidates $(1,1)$ and $(1,2)$ survive with probability close to one and $(1,3)$ in about $75\%$ of cases, whereas $(1,4)$ survives in fewer than $40\%$. SCP nevertheless keeps $(1,4)$ and skips better candidates, which is the optimal child-node skip problem and the main source of its BER loss.
\end{enumerate}
Among the remaining candidates, $(2,1)$ has the highest survival probability. Its survival trend complements that of $(1,4)$ and dominates at low SNR, where $(1,4)$ is least likely to survive. MMECP exploits this complementary behavior.

\subsection{MMECP Principle}

MMECP starts from the SCP survivor set $\mathcal G^{(1)}_i$ and changes at most one of its $K$ paths. By observation~2 in Section~\ref{sec:surv_analysis}, the SCP survivor most likely to be wrong is $(1,K)$, the candidate with the largest accumulated PED in $\mathcal G^{(1)}_i$, and the strongest replacement is $(2,1)$, the candidate with the smallest accumulated PED in $\mathcal G^{(2)}_i$. Let $\hat W^{(1)}_{\max,i}$ and $\hat W^{(2)}_{\min,i}$ denote their paths. MMECP keeps the one with the smaller accumulated PED and discards the other. We call this step the \emph{max-min exchange}. The rule
\begin{enumerate}
    \item targets $(1,K)$, which survives global sorting in fewer than $40\%$ of cases;
    \item uses $(2,1)$, whose survival trend complements that of $(1,K)$; and
    \item requires no cross-comparisons with other candidates, whose survival probabilities are negligible.
\end{enumerate}
The rule uses only the ordering within each Box cluster from Rule~2 and the PEDs of the rank groups $\mathcal G^{(1)}_i$ and $\mathcal G^{(2)}_i$. It therefore applies to any detector whose children are ordered per parent, including the on-demand $K$-Best of~\cite{DKB_v1}. Reduced sorters for $K$-Best~\cite{k_best_sort_free, KBest_Sorter_2017, Sorter_2023} still compare the candidates of different parents at every layer. MMECP needs only the max tree, the min tree and one final comparison across parents, because Rules~1 and~2 order the candidates of each parent without any PED comparison.

\emph{Box-SMMECP.} Following SICP, MMECP is applied only at the first pruning layer and SCP at the remaining layers. The resulting detector, Box-SMMECP, is the iteration-free counterpart of Box-SICP and is implemented together with Box-SCP in Section~\ref{sec:arch}.

\begin{figure}[t]
        \centering
        \includegraphics[width=0.40\textwidth]{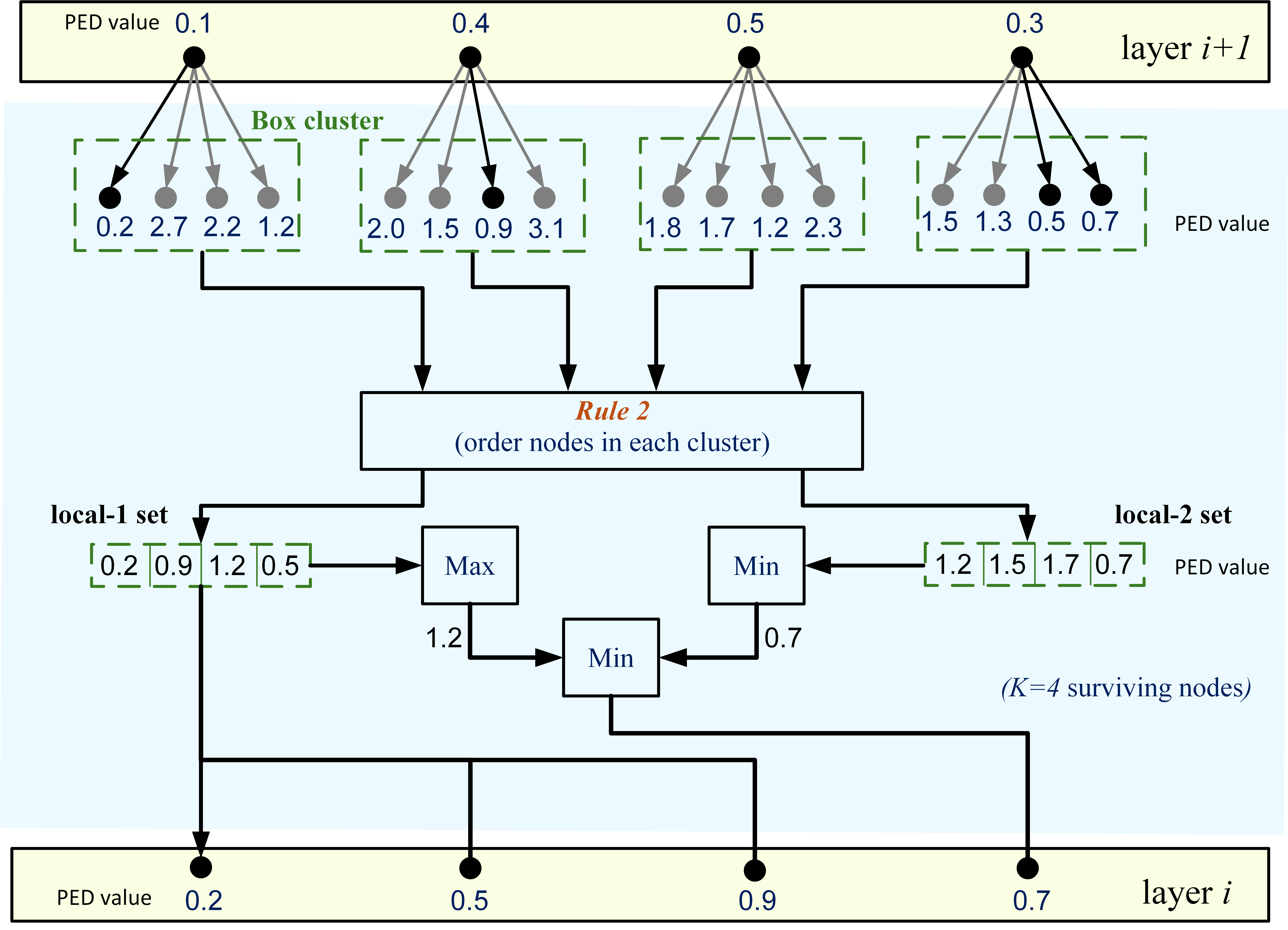}
        \caption{MMECP block diagram at layer $i$ ($K=4$).}
        \label{fig:Box_MMECP}
\end{figure}

Fig.~\ref{fig:Box_MMECP} shows the block-level structure of MMECP at layer~$i$ for $K=4$. Algorithm~\ref{algo: MMECP} summarizes the candidate expansion and selection with MMECP in each layer.

\begin{algorithm}[t]
\caption{Candidate Expansion and Selection with MMECP}
\begin{algorithmic}[1]

\REQUIRE  $\bh{\mathbb{X}}_{i+1} = \{(\bh{X}_{k,i+1}, \bh{d}_{k,i+1})\}_{k=1}^{K}$ with $\bh{X}_{k,i+1} =\{\hat{x}_{k,j}\}_{j=i+1}^N$, and the reference nodes $\{\hat a_{k,i}\}_{k=1}^{K}$ from~\eqref{eq:box_ai_alg}
\ENSURE $\bh{\mathbb{X}}_{i}$

\STATE Initialize the augmented rank groups: $\bh{\mathbb W}_{i}^{(1)}, \bh{\mathbb W}_{i}^{(2)} \gets \emptyset$.

\FOR{$k = 1$ to $K$}
    \STATE $\hat{\mathcal B}_{k,i} = \{\hat{x}_{k,i}(l)\}_{l=1}^{B} \gets \mathrm{Box}(\hat a_{k,i})$
    \STATE $\hat{x}^{(1)}_{k,i} \gets \mathrm{Rule1}(\hat{\mathcal B}_{k,i},\hat a_{k,i})$
    \STATE $\hat{x}^{(2)}_{k,i} \gets \mathrm{Rule2}(\hat{\mathcal B}_{k,i},\hat a_{k,i},\hat{x}^{(1)}_{k,i})$
    \STATE $\bh{W}^{(1)}_{k,i}\gets(\bh{X}_{k,i+1},\hat{x}_{k,i}^{(1)})$, $\bh{d}^{(1)}_{k,i} \gets \mathrm{PED}(\bh{W}^{(1)}_{k,i})$
    \STATE $\bh{W}^{(2)}_{k,i}\gets(\bh{X}_{k,i+1},\hat{x}_{k,i}^{(2)})$, $\bh{d}^{(2)}_{k,i} \gets \mathrm{PED}(\bh{W}^{(2)}_{k,i})$
    \STATE $\bh{\mathbb W}_{i}^{(1)} \gets \bh{\mathbb W}_{i}^{(1)} \cup \{(\bh{W}^{(1)}_{k,i}, \bh{d}^{(1)}_{k,i})\}$
    \STATE $\bh{\mathbb W}_{i}^{(2)} \gets \bh{\mathbb W}_{i}^{(2)} \cup \{(\bh{W}^{(2)}_{k,i}, \bh{d}^{(2)}_{k,i})\}$
\ENDFOR

\STATE $k_{\max} \gets \arg\max\limits_{k\in\{1,\dots,K\}} \bh{d}^{(1)}_{k,i}$, $k_{\min} \gets \arg\min\limits_{k\in\{1,\dots,K\}} \bh{d}^{(2)}_{k,i}$
\STATE $(\bh{W}^{(1)}_{\max,i},\bh{d}^{(1)}_{\max,i}) \gets (\bh{W}^{(1)}_{k_{\max},i},\bh{d}^{(1)}_{k_{\max},i})$,\newline $(\bh{W}^{(2)}_{\min,i},\bh{d}^{(2)}_{\min,i}) \gets (\bh{W}^{(2)}_{k_{\min},i},\bh{d}^{(2)}_{k_{\min},i})$

\STATE $\bh{\mathbb X}_{i} \gets \bh{\mathbb W}_{i}^{(1)} \setminus \{(\bh{W}_{\max,i}^{(1)}, \bh{d}^{(1)}_{\max,i})\}$

\IF{$\bh{d}_{\min,i}^{(2)} < \bh{d}_{\max,i}^{(1)}$}
    \STATE $\bh{\mathbb X}_{i} \gets \bh{\mathbb X}_{i} \cup \{(\bh{W}_{\min,i}^{(2)}, \bh{d}_{\min,i}^{(2)})\}$
\ELSE
    \STATE $\bh{\mathbb X}_{i} \gets \bh{\mathbb X}_{i} \cup \{(\bh{W}_{\max,i}^{(1)},\bh{d}_{\max,i}^{(1)})\}$
\ENDIF

\STATE \textbf{return} $\bh{\mathbb X}_{i}$

\end{algorithmic}
\label{algo: MMECP}
\end{algorithm}

\subsection{Sorting Accuracy}
\label{sec:sort_acc}
To show how MMECP changes the selected survivors, Fig.~\ref{fig:agreement_curves} compares SCP, MMECP, and ICP in terms of \emph{sorting accuracy}, defined as the fraction of the $K$ survivors of a layer that match those of global sorting, averaged over all layers and channel realizations. The \emph{layer-wise} accuracy in Fig.~\ref{fig:ll_agr} takes the global-sorting survivors from the same parent paths as the pruning under test, so it measures the decision of a single layer. The \emph{tree-wise} accuracy in Fig.~\ref{fig:tl_agr} takes the survivors of a detector that applies global sorting at every layer, so it also includes the deviations propagated from the layers above. In Fig.~\ref{fig:ll_agr}, MMECP exceeds $80\%$ across the SNR range, $10$ to $20$ percentage points above SCP, while ICP performs exact sorting and reaches $100\%$.

In Fig.~\ref{fig:tl_agr}, \mbox{Box-SMMECP} reaches a tree-wise accuracy of $70$ to $80\%$, compared with $55$ to $70\%$ for \mbox{Box-SCP}, and is close to \mbox{Box-SICP}. Section~\ref{sec:Sim} examines how these differences affect the BER.

\section{Numerical Results}
\label{sec:Sim}
\subsection{Computational Complexity}
\label{sec:complexity}
Table~\ref{tab:prune_cost} lists the real floating-point operations (FLOPs) of one pruning layer, counted per step as in~\cite{box-scp}. The Box operation and the interference cancellation are the same for all schemes and are not included. ICP evaluates the PEDs of the $K$ first winners and then, in each round, the PED of the next candidate of the Box cluster just selected, i.e., $K+(K-1)=2K-1$ PEDs in total. Which candidate is evaluated is known only after the previous round, so a PED unit sits inside every selection round. The $K$ dependent rounds can be \emph{folded} into one loop reused over $K$ cycles, which saves area but accepts a new layer input only every $K$ cycles, or \emph{unfolded} into $U$ individual round units with $2\le U\le K$, which accept one input every $\lceil K/U\rceil$ cycles at about $U$ times the round hardware. ICP therefore faces the trade-off that any iteration mapped to hardware has. It trades area against throughput. MMECP costs about the same number of FLOPs as ICP but accepts one input per cycle without a selection loop. Its comparisons form two parallel trees and one final comparator, i.e., $\lceil\log_2K\rceil+1$ stages. The $K$ dependent rounds of ICP need $K\lceil\log_2K\rceil$ stages and $K(K-1)$ comparisons, which grow with $K^2$ instead of $K$. Since Box-SMMECP applies MMECP only at the first pruning layer, its pruning cost exceeds that of Box-SCP by only $47$ FLOPs per detected vector for $K=4$.

\begin{table}[t]
\caption{Pruning cost of one detection layer in real FLOPs with $K$ survivors from the last layer ($K=4$).}
\label{tab:prune_cost}
\centering
\footnotesize
\setlength{\tabcolsep}{4pt}
\begin{tabular}{@{}lccc@{}}
\toprule
 & SCP & MMECP & ICP\\
\midrule
Rule~1 & $2K$ & $2K$ & $2K$\\
Rule~2 & -- & $3K$ & $3K$\\
PED evaluation & $7K$ & $14K$ & $7(2K-1)$\\
Selection comparison & -- & $2K-1$ & $K(K-1)$\\
\midrule
Total & $36$ & $83$ & $81$\\
Comparator stages & $0$ & $3$ & $8$\\
Cycles per layer input & $1$ & $1$ & $4$--$1$$^{\mathrm{a}}$\\
\bottomrule
\end{tabular}\\[2pt]
{\scriptsize $^{\mathrm{a}}$$4$ with one folded selection loop, $1$ with $K$ unfolded round units.}
\end{table}

\subsection{BER Performance}
\label{sec:ber}
\begin{figure*}[t]
    \centering
    \begin{subfigure}[t]{0.193\textwidth}
        \centering
        \includegraphics[width=\linewidth]{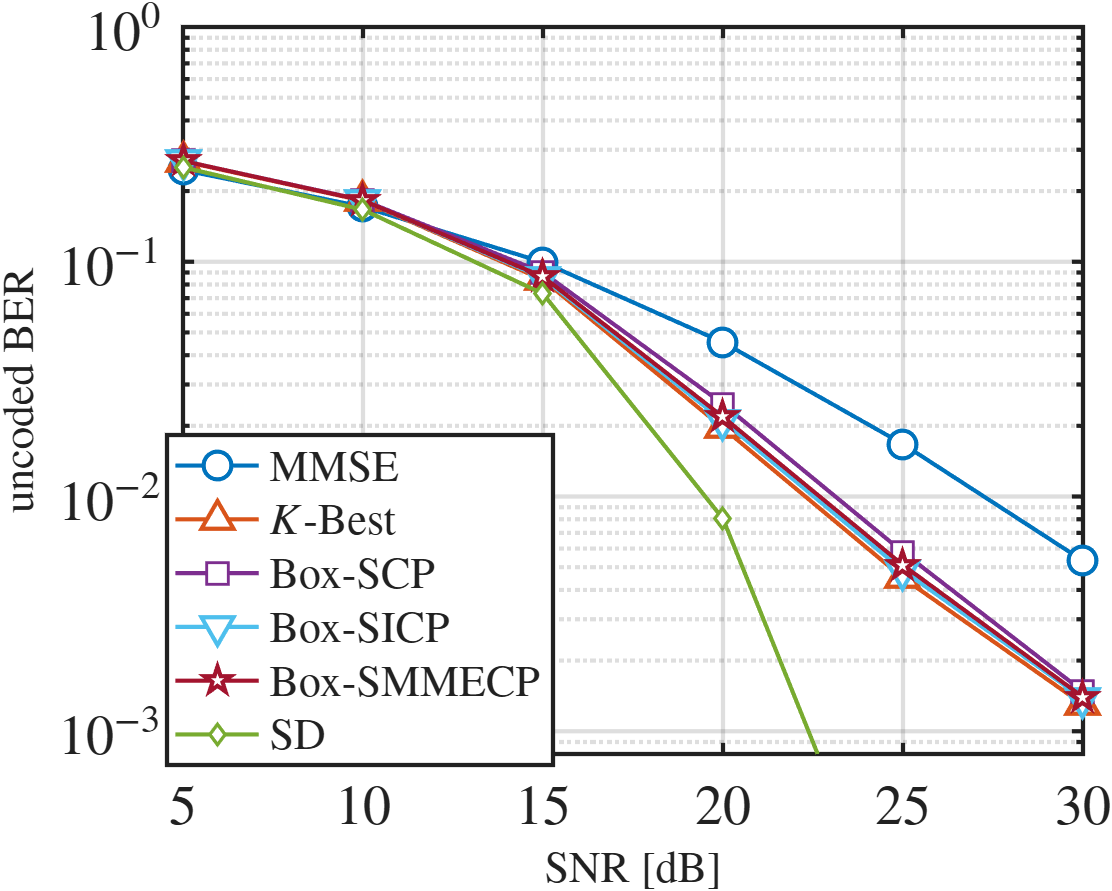}
        \caption{$64$-QAM}
        \label{fig:ber64}
    \end{subfigure}
    \hfill
    \begin{subfigure}[t]{0.193\textwidth}
        \centering
        \includegraphics[width=\linewidth]{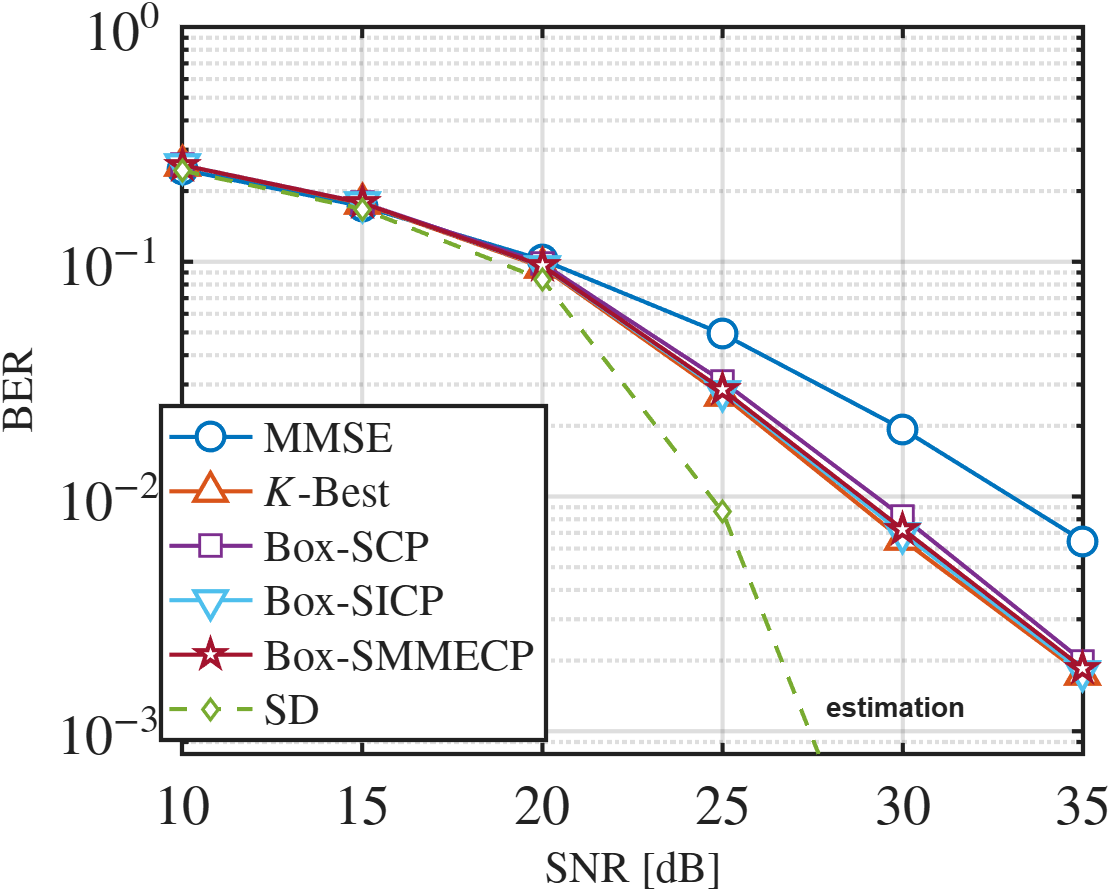}
        \caption{$256$-QAM}
        \label{fig:ber256}
    \end{subfigure}
    \hfill
    \begin{subfigure}[t]{0.193\textwidth}
        \centering
        \includegraphics[width=\linewidth]{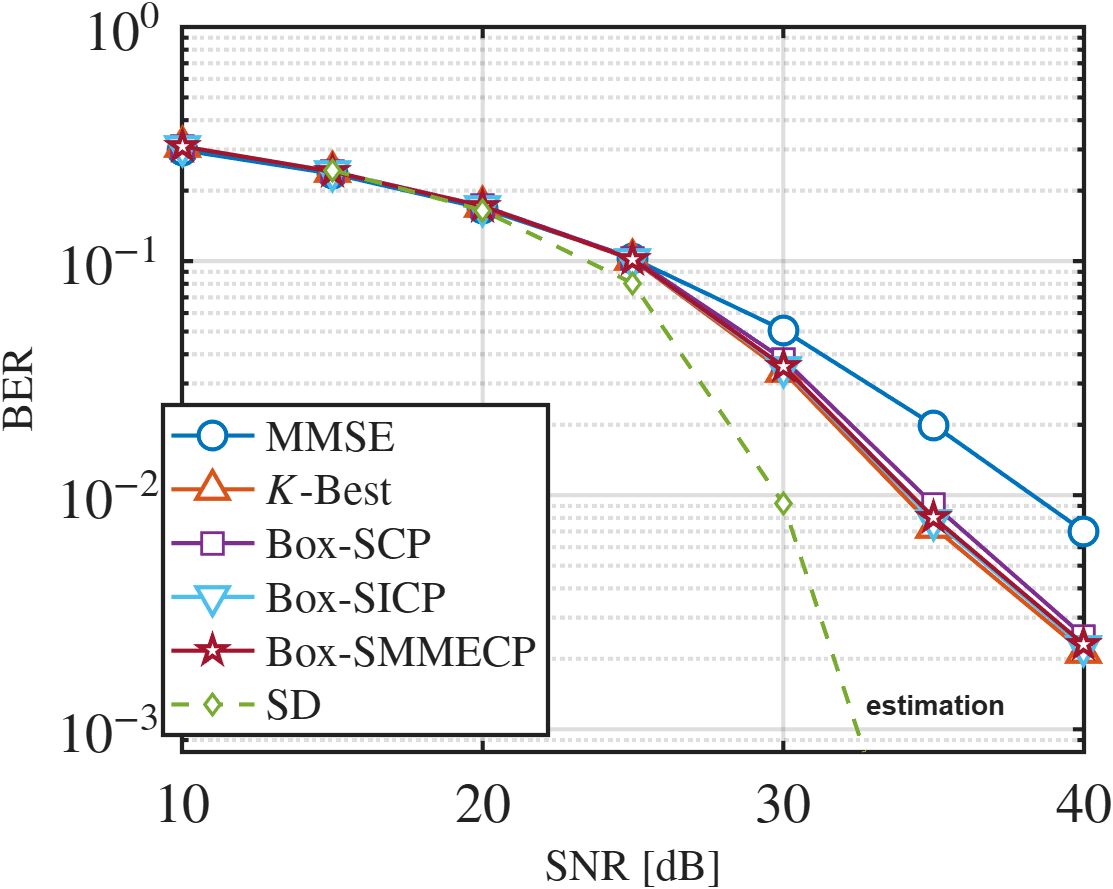}
        \caption{$1024$-QAM}
        \label{fig:ber1024}
    \end{subfigure}
    \hfill
    \begin{subfigure}[t]{0.193\textwidth}
        \centering
        \includegraphics[width=\linewidth]{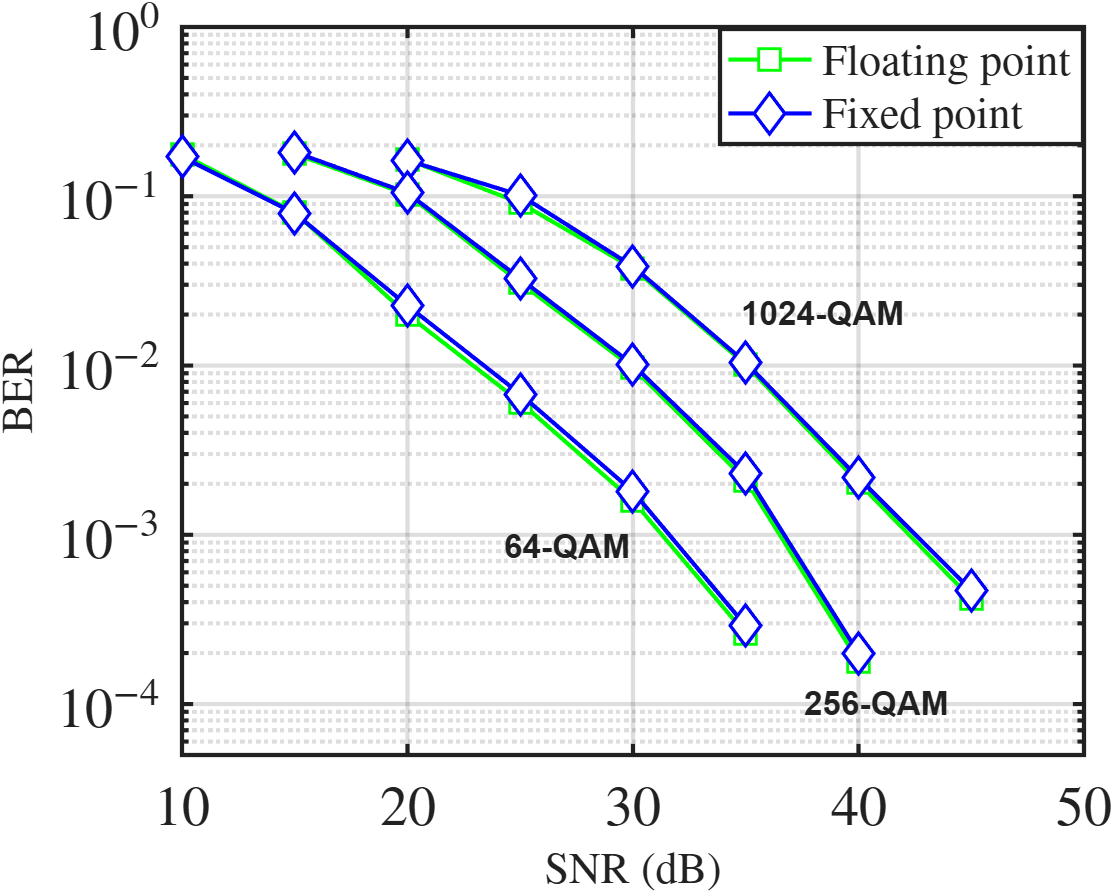}
        \caption{Box-SCP, co-sim.}
        \label{fig:cosim_ber_scp}
    \end{subfigure}
    \hfill
    \begin{subfigure}[t]{0.193\textwidth}
        \centering
        \includegraphics[width=\linewidth]{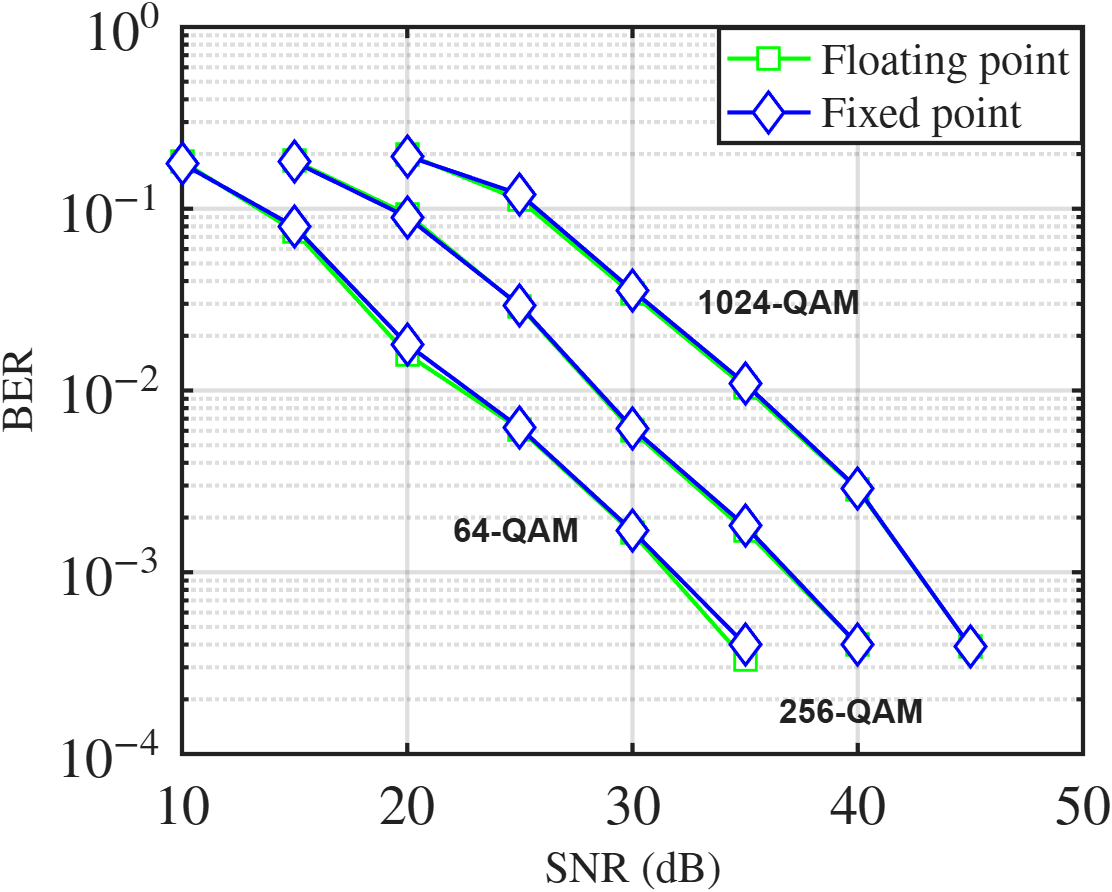}
        \caption{Box-SMMECP, co-sim.}
        \label{fig:cosim_ber_smcp}
    \end{subfigure}
    \caption{Uncoded BER of the $4\times4$ MIMO detector over an i.i.d.\ Rayleigh channel at $64$-, $256$-, and $1024$-QAM. (a)--(c) Floating-point BER of all detectors. (d), (e) Fixed-point versus floating-point BER.}
    \label{fig:ber_all}
\end{figure*}

This subsection evaluates the BER performance of the proposed \mbox{Box-SMMECP} detector against \mbox{Box-SCP}, \mbox{Box-SICP}, and $K$-Best ($K=4$). Linear MMSE and sphere decoding (SD) are included as the linear and near-ML references. \mbox{$K$-Box} and DKB are not plotted separately, since they attain the same detection performance as \mbox{$K$-Best} for the same $K$~\cite{KBox, DKB_conf}. The \mbox{Box-ICP} and ZF results are already reported in~\cite{box-scp} and are not repeated. The number of surviving nodes per layer is fixed at $K=B=4$ for all $K$-Best and Box schemes. All results use a $4\times4$ MIMO detector over an uncoded i.i.d.\ Rayleigh flat-fading channel with perfect channel state information (CSI). The modulation order ranges from $64$- to $1024$-QAM (Figs.~\ref{fig:ber64}--\ref{fig:ber1024}).

SD serves as the near-ML reference. At $64$-QAM its BER is simulated directly. At $256$- and $1024$-QAM an exhaustive SD search is impractical, so the SD curve is estimated by shifting the lower-order curve along the SNR axis. For QAM over AWGN the error probability depends on SNR only through $E_s/((|\mathcal{A}|-1)N_0)$~\cite{goldsmith2005wireless}, so each fourfold increase in constellation size raises the required SNR by about $10\log_{10}4\approx6$~dB at a fixed BER, with the diversity slope unchanged. Each step ($64\!\to\!256\!\to\!1024$) therefore shifts the SD curve right by about $6$~dB, which gives the $256$- and $1024$-QAM references.

Figs.~\ref{fig:ber64}--\ref{fig:ber1024} show the uncoded BER for $64$-, $256$-, and $1024$-QAM. The ordering of the curves is the same at every modulation order. SD gives the best performance and MMSE the worst, with the $K$-Best and Box detectors in between. Among the tree-search schemes, \mbox{Box-SCP} shows the largest loss, caused by the optimal child-node skip problem of single-step pruning. It trails $K$-Best by about $0.8$~dB from $64$- to $1024$-QAM at a BER of $10^{-2}$. \mbox{Box-SMMECP} improves on \mbox{Box-SCP} at every modulation order and almost overlaps \mbox{Box-SICP}, narrowing the gap to $K$-Best to about $0.2$~dB. This agrees with the tree-wise sorting accuracy in Fig.~\ref{fig:tl_agr}, where \mbox{Box-SMMECP} is close to \mbox{Box-SICP}. The single max-min exchange of MMECP therefore improves the detection performance and keeps the fully parallel, feed-forward structure of SCP.

\subsection{Fixed-Point Quantization}
\label{sec:cosim}
The detector is implemented in fixed-point arithmetic. We assume that the hardware datapath operates on the energy-unnormalized integer QAM grid with $\Delta q=2$, as in other $K$-Best designs~\cite{k_best_sort_free,DKB_v1,sort_free_2024}, so $\gamma=-M$ (Section~\ref{sec:system_model}) and the shift between $\mathcal{A}$ and the index QAM map $\tilde{\mathcal{A}}$ (Section~\ref{sec:idx_map}) is written with $M$. The word length of each variable is written in the Q$m.n$ format, with $m$ integer bits including the sign and $n$ fractional bits. The integer bits follow from the dynamic range of each variable in simulation with a Rayleigh fading channel and random transmit symbols. The fractional bits are the fewest that keep the BER within about $0.1$\,dB of double-precision floating point at a BER of $10^{-3}$. The fixed constants, namely the QAM spacing $\Delta q=2$, its reciprocal $0.5$, and the offset $M$ of $\tilde{\mathcal A}$, use minimal word lengths. Most datapath variables, including the received signal $y'$, the entries of $\mathbf{R}$, and the PED, share a common word length of Q$7.7$, Q$8.8$, and Q$9.9$ for $64$-, $256$-, and $1024$-QAM, respectively.

The main exception is the candidate index $\tilde x$ in the index QAM map $\tilde{\mathcal{A}}$, which is integer-only and sized by the QAM order. The candidate index $\tilde x(l)$ spans $[0,M]$, so its field needs $\lceil\log_2(M{+}1)\rceil$ bits plus one guard bit for a reference node $\tilde a$ that sits below $0$ or above $M$ before clipping, giving $4$, $5$, and $6$ bits for $64$-, $256$-, and $1024$-QAM ($M{=}7,15,31$). This width grows by one bit per QAM-order step, set by the index range rather than the symbol-domain amplitude of $\hat a$ in $\mathcal A$.

Figs.~\ref{fig:cosim_ber_scp} and~\ref{fig:cosim_ber_smcp} validate the fixed-point design by hardware co-simulation, which compares the BER of the bit-accurate hardware model with that of the floating-point algorithm on the same transmissions. The hardware model uses the $\ell_1$-norm PED~\eqref{eq:l1_ped} of Section~\ref{sec:idx_ped}, whereas the floating-point reference uses the squared Euclidean PED~\eqref{eqn:PEDa}, so the comparison also covers the change of norm. Across $64$-, $256$- and $1024$-QAM, the fixed-point BER curves overlap the floating-point reference over the full SNR range, with no visible degradation, so the chosen word lengths are sufficient.


\section{Index-Domain Reformulation}
\label{sec:idx_alg}
The algorithm in Sections~\ref{sec:system_model} and~\ref{sec:pruning} defines the detection in the QAM map $\mathcal{A}$. This section reformulates the same detection on the integer indices of the constellation points. The reformulation does not change the detected symbols. It replaces operations on symbol values by operations on index bits, so that every detection step has a short and fixed computation. The fully pipelined architecture of Section~\ref{sec:arch} relies on this property. Within one Box cluster at one layer, the subscript $_{k,i}$ is dropped where no confusion arises, so $\tilde a$ stands for $\tilde a_{k,i}$ and $\tilde\delta(l)$ for $\tilde\delta_{k,i}(l)$.

\subsection{Index QAM Map $\tilde{\mathcal{A}}$}
\label{sec:idx_map}
In the QAM map $\mathcal{A}$, each quantity carries a hat, such as $\hat a$, $\hat\delta$, and $\hat x$. The reformulated detection instead operates in the index QAM map $\tilde{\mathcal{A}}$, where the corresponding quantities carry a tilde, such as $\tilde a$, $\tilde\delta$, and $\tilde x$. \par

Section~\ref{sec:system_model} describes each QAM grid point by its grid indices $n$ and $p$. The index QAM map collects these indices,
\begin{equation}
\tilde{\mathcal{A}} = \{\, n+\mathrm{j}p \;|\; n,p\in\{0,\dots,M\} \,\},
\label{eq:idx_map}
\end{equation}
and a symbol $\hat x\in\mathcal{A}$ and its index $\tilde x\in\tilde{\mathcal{A}}$ are related by
\begin{equation}
\hat x = \Delta q\,\tilde x - M(1+\mathrm{j}),
\label{eq:idx_sym}
\end{equation}
with $\gamma=-M$ on the integer grid of Section~\ref{sec:cosim}. For $4$-QAM, for example, $M=1$, $\mathcal{A}=\{-1,1\}\times\{-1,1\}$, and $\tilde{\mathcal{A}}=\{0,1\}\times\{0,1\}$. The reference node maps to $\tilde a=(\hat a+M(1+\mathrm{j}))/\Delta q$, which is exactly the argument of the floor in~\eqref{eqn:box_operation}. The Box operation therefore already works on indices. \par

A direct evaluation of~\eqref{eqn:box_operation} performs an index map--demap round-trip $\mathcal{A}\!\to\!\tilde{\mathcal{A}}\!\to\!\mathcal{A}$ per Box cluster. The reference node $\hat a$ is mapped to $\tilde a$ to enable the floor. The resulting index $\tilde x(l)$ is demapped to $\hat x(l)\in\mathcal{A}$ with~\eqref{eq:idx_sym}, and the PED is accumulated from $\hat\delta(l) = \hat a - \hat x(l)$ in $\mathcal{A}$. The offset by $M$ is shared by both $\hat x(l)$ and $\hat a$ and cancels in $\hat\delta(l)$. The proposed reformulation removes the demapping by keeping all steps after the floor in $\tilde{\mathcal{A}}$ and accumulating an equivalent normalized PED~\eqref{eq:idx_ped} directly from the \emph{$\Delta q$-normalized displacement} $\tilde\delta(l) = \tilde a - \tilde x(l)$, whose real and imaginary parts are its distance components. \par

The integer part $\mathfrak{i}$ and the fractional part $\mathfrak{f}$ of $\tilde a$, extracted by the operators $\mathfrak{int}(\cdot)$ and $\mathfrak{frac}(\cdot)$, have two separate roles. The integer part $\mathfrak{i}$ is reused as the starting candidate index $\tilde x(1)$, and the fractional part $\mathfrak{f}$ gives the displacement $\tilde\delta(l)$ of each candidate in closed form. Both are derived below. The real and imaginary components of $\tilde a$ are treated independently. The real component is used as the example below, and the imaginary component follows the same rules. \par

\paragraph{Candidate index assignment}
In the real dimension, the floor gives the index of the starting candidate, $\mathfrak{R}\{\tilde x(1)\}=\lfloor\mathfrak{R}\{\tilde a\}\rfloor$. With noise, $\mathfrak{R}\{\tilde a\}$ may fall outside the range $[0, M]$ of $\tilde{\mathcal{A}}$, so three regions are distinguished, \emph{neg} ($\mathfrak{R}\{\tilde a\} \leq 0$), \emph{mid} ($0 < \mathfrak{R}\{\tilde a\} < M$), and \emph{ovf} ($\mathfrak{R}\{\tilde a\} \geq M$). The starting candidate index $\tilde x(1)$ takes different values according to the region of $\tilde a$. $\mathfrak{R}\{\tilde x(1)\}$ is
\begin{enumerate}
    \item set to the integer part $\mathfrak{i}^{\mathrm{re}}=\mathfrak{int}(\mathfrak{R}\{\tilde a\})$ in \emph{mid},
    \item clipped to the lower boundary $0$ in \emph{neg}, and
    \item clipped to the upper boundary $M-1$ in \emph{ovf}.
\end{enumerate}

The imaginary component follows the same rule, so $\tilde a$ falls into one of nine regions of the constellation. Table~\ref{tab:int_assign} lists $\tilde x(1)$ for the nine regions.\par
Owing to the Box geometry, the other three candidate indices inside one Box cluster in $\tilde{\mathcal{A}}$ differ from $\tilde x(1)$ by fixed integer offsets,
\begin{equation}
\tilde x(l) = \tilde x(1)+ o_l, \quad o_2=\mathrm{j},\ o_3=1,\ o_4=1+\mathrm{j}.
\label{eq:idx_offset}
\end{equation}

\paragraph{Normalized displacement assignment}
Substituting Table~\ref{tab:int_assign} and~\eqref{eq:idx_offset} into the displacement $\tilde\delta(l) = \tilde a - \tilde x(l)$ yields a closed form per region. Taking $\mathfrak{R}\{\tilde\delta(1)\}$ as an example, it equals
\begin{enumerate}
    \item the fractional part $\mathfrak{f}^{\mathrm{re}}=\mathfrak{frac}(\mathfrak{R}\{\tilde a\})$ in \emph{mid};
    \item $\mathfrak{R}\{\tilde a\}$ itself in \emph{neg};
    \item    $\mathfrak{R}\{\tilde a\}-(M-1)$ in \emph{ovf}.
\end{enumerate}
When $\mathfrak{R}\{\tilde a\}$ falls outside $[0,M]$, the clipped Box still contains the constellation point nearest to $\tilde a$, so the clipping is exact at the boundary and loses no detection accuracy. Table~\ref{tab:dis_comp} lists $\tilde\delta(1)$ for the nine regions. The displacements $\tilde\delta(l)$ of the other candidates for $l=2,3,4$ are obtained from $\tilde\delta(1)$ by subtracting the Box offset $o_l$ of~\eqref{eq:idx_offset} in the real or imaginary dimension. Absolute-value signs are omitted, since the $\ell_1$-norm PED~\eqref{eq:l1_ped} is used for the implementation and it only takes the magnitude (Section~\ref{sec:idx_ped}). \par

\begin{table*}[!t]
\caption{Closed-form value assignment per region of $\tilde a$, in $(\mathfrak{R},\mathfrak{I})$ form.}
\label{tab:An_assign}
\centering
\begin{subtable}{0.49\textwidth}
\centering
\caption{Starting candidate index $\tilde x(1)$}
\label{tab:int_assign}
\footnotesize
\begin{tabular}{@{}lccc@{}}
\toprule
 & \multicolumn{3}{c}{$\mathfrak{R}\{\tilde a\}$ position}\\
\cmidrule(lr){2-4}
$\mathfrak{I}\{\tilde a\}$ pos. & mid & neg & ovf\\
\midrule
mid & $(\mathfrak{i}^{\mathrm{re}},\,\mathfrak{i}^{\mathrm{im}})$ & $(0,\,\mathfrak{i}^{\mathrm{im}})$ & $(M{-}1,\,\mathfrak{i}^{\mathrm{im}})$\\
neg & $(\mathfrak{i}^{\mathrm{re}},\,0)$ & $(0,\,0)$ & $(M{-}1,\,0)$\\
ovf & $(\mathfrak{i}^{\mathrm{re}},\,M{-}1)$ & $(0,\,M{-}1)$ & $(M{-}1,\,M{-}1)$\\
\bottomrule
\end{tabular}\\[2pt]
{\footnotesize $\mathfrak{i}^{\mathrm{re}},\mathfrak{i}^{\mathrm{im}}$: integer parts of $\tilde a$ in real / imag.\ dim.}
\end{subtable}
\hfill
\begin{subtable}{0.49\textwidth}
\centering
\caption{Normalized displacement $\tilde\delta(1)$}
\label{tab:dis_comp}
\footnotesize
\begin{tabular}{@{}lccc@{}}
\toprule
 & \multicolumn{3}{c}{$\mathfrak{R}\{\tilde a\}$ position}\\
\cmidrule(lr){2-4}
$\mathfrak{I}\{\tilde a\}$ pos. & mid & neg & ovf\\
\midrule
mid & $(\mathfrak{f}^{\mathrm{re}},\,\mathfrak{f}^{\mathrm{im}})$ & $(\mathfrak{R}\{\tilde a\},\,\mathfrak{f}^{\mathrm{im}})$ & $(\omega^{\mathrm{re}}{+}1,\,\mathfrak{f}^{\mathrm{im}})$\\
neg & $(\mathfrak{f}^{\mathrm{re}},\,\mathfrak{I}\{\tilde a\})$ & $(\mathfrak{R}\{\tilde a\},\,\mathfrak{I}\{\tilde a\})$ & $(\omega^{\mathrm{re}}{+}1,\,\mathfrak{I}\{\tilde a\})$\\
ovf & $(\mathfrak{f}^{\mathrm{re}},\,\omega^{\mathrm{im}}{+}1)$ & $(\mathfrak{R}\{\tilde a\},\,\omega^{\mathrm{im}}{+}1)$ & $(\omega^{\mathrm{re}}{+}1,\,\omega^{\mathrm{im}}{+}1)$\\
\bottomrule
\end{tabular}\\[2pt]
{\footnotesize $\mathfrak{f}^{\mathrm{re}},\mathfrak{f}^{\mathrm{im}}$: fractional parts of $\tilde a$; $\omega^{\mathrm{re}}\!\equiv\!\mathfrak{R}\{\tilde a\}\!-\!M$, $\omega^{\mathrm{im}}\!\equiv\!\mathfrak{I}\{\tilde a\}\!-\!M$.}
\end{subtable}
\end{table*}

\subsection{Reference Node in $\tilde{\mathcal{A}}$}
\label{sec:idx_rng}
We rewrite~\eqref{eq:box_ai_alg} in $\tilde{\mathcal{A}}$ as
\begin{equation}
\begin{aligned}
\tilde a &= \frac{\hat a + M(1+\mathrm{j})}{\Delta q} \\
&= \frac{y'_i}{r_{i,i}\Delta q} - \sum_{j=i+1}^{N} \rho_{i,j}\,\hat x_j +\lambda,
\end{aligned}
\label{eq:idx_a}
\end{equation}
where
\begin{equation}
  \rho_{i,j} = \frac{r_{i,j}}{r_{i,i}\Delta q},\quad  \lambda = \frac{M(1+\mathrm{j})}{\Delta q}.
\end{equation}
The \textit{interference coefficient} $\rho_{i,j}$ and the scaling $1/(r_{i,i}\Delta q)$ depend only on the channel matrix, so they can be pre-computed once the QR decomposition is completed. \par

The symbols detected at the previous layers are kept as indices $\tilde x_j \in \tilde{\mathcal{A}}$ rather than as QAM symbols $\hat x_j$\footnote{The imaginary unit is typeset upright as $\mathrm{j}=\sqrt{-1}$ to distinguish it from the summation index $j$.}. Substituting~\eqref{eq:idx_sym} into the interference term of~\eqref{eq:idx_a} gives
\begin{equation}
\sum_{j=i+1}^{N} \rho_{i,j}\,\hat x_j
= \Delta q \sum_{j=i+1}^{N} \rho_{i,j}\,\tilde x_j - \sum_{j=i+1}^{N} \rho_{i,j}\,M(1+\mathrm{j}) ,
\label{eq:idx_interf}
\end{equation}
so each multiplication becomes a product of $\rho_{i,j}$ and an integer index $\tilde x_j$, whose real and imaginary parts lie in $\{0,\dots,M\}$. The constant term $\sum_{j=i+1}^{N}\rho_{i,j}\,M(1+\mathrm{j})$ of~\eqref{eq:idx_interf} does not depend on the path. \par

\subsection{Rule Reformulation in $\tilde{\mathcal{A}}$}
\label{sec:idx_rule}
Substituting $\hat\delta = \Delta q\,\tilde\delta$ into the original definitions~\eqref{eqn: Rule1} and~\eqref{eqn: Rule2} cancels the $\Delta q$ factor and gives
\begin{equation}
\begin{aligned}
    &1 - 2\mathfrak{R}\{\tilde\delta(1)\} & \gtrless0,&\\
    &1 - 2\mathfrak{I}\{\tilde\delta(1)\} & \gtrless0,&\\
    &\mathfrak{R}\{\tilde\delta(1)\}-\mathfrak{I}\{\tilde\delta(1)\}   & \gtrless0,&\\
    &\mathfrak{R}\{\tilde\delta(1)\}+\mathfrak{I}\{\tilde\delta(1)\}-1 & \gtrless0.&
\end{aligned}
\label{eq:idx_rules}
\end{equation}
The first two inequalities form Rule~1 and the latter two form Rule~2. The sign of each inequality decides which of the two candidates it compares is closer to $\tilde a$. \par

\subsection{Normalized PED}
\label{sec:idx_ped}
With~\eqref{eq:box_ai_alg}, the term inside $|\cdot|$ in~\eqref{eqn:PEDa} equals $r_{i,i}\hat\delta(l)$ for a real $r_{i,i}$. Substituting $\hat\delta(l) = \Delta q\,\tilde\delta(l)$ into~\eqref{eqn:PEDa}, the path PED after appending the $l$-th candidate at layer $i$ is
\begin{equation}
\bh{d}_{k,i}(l) = \bh{d}_{k,i+1} + r_{i,i}^2\,|\hat\delta(l)|^2 = \bh{d}_{k,i+1} + (r_{i,i}\Delta q)^2\,|\tilde\delta(l)|^2.
\label{eq:idx_ped_A}
\end{equation}
Factoring $(\Delta q)^2$ out of every per-layer increment defines the $(\Delta q)^2$-normalized PED $\tilde d_{k,i}(l) = \bh{d}_{k,i}(l)/(\Delta q)^2$, which accumulates as
\begin{equation}
\tilde d_{k,i}(l) = \tilde d_{k,i+1} + r_{i,i}^2\,|\tilde\delta(l)|^2,
\label{eq:idx_ped}
\end{equation}
which depends only on $\tilde\delta(l)$ and the per-layer constant $r_{i,i}$. Since the $1/(\Delta q)^2$ factor is the same across all paths and layers, accumulating $\tilde d$ in place of $\bh{d}$ preserves the survivor ordering and avoids the $(\Delta q)^2$ scaling on every increment. Written out in its components,~\eqref{eq:idx_ped} is
\begin{equation*}
\tilde d_{k,i}(l) = \tilde d_{k,i+1} + r_{i,i}^2\bigl(\mathfrak{R}\{\tilde\delta(l)\}^2+\mathfrak{I}\{\tilde\delta(l)\}^2\bigr).
\end{equation*}
To remove the squaring in hardware, the squared Euclidean norm $|r_{i,i}\tilde\delta(l)|^2$ is replaced by the $\ell_1$-norm $\|r_{i,i}\tilde\delta(l)\|_1$, as in~\cite{multigigabit_2014}. With $r_{i,i}>0$, this gives the $\ell_1$-norm PED
\begin{equation}
\begin{aligned}
\tilde d_{k,i}(l) &= \tilde d_{k,i+1} + r_{i,i}\,\|\tilde\delta(l)\|_1\\
&= \tilde d_{k,i+1} + r_{i,i}\bigl(|\mathfrak{R}\{\tilde\delta(l)\}|+|\mathfrak{I}\{\tilde\delta(l)\}|\bigr).
\end{aligned}
\label{eq:l1_ped}
\end{equation}
Rule~1 and Rule~2 stay the same, since each of their comparisons reduces to the same linear inequality for both norms. \par

\subsection{Hardware Advantages of the $\tilde{\mathcal{A}}$ Datapath}
\label{sec:idx_hw}
The demapping of Section~\ref{sec:idx_map} costs a scaling by $\Delta q$ and a subtraction per candidate coordinate. The closed-form assignments in Table~\ref{tab:An_assign} replace the demapping with bit selection on $\tilde a$ and the two region tests. In the \emph{mid} region, $\mathfrak{R}\{\tilde\delta(1)\}=\mathfrak{f}^{\mathrm{re}}$, so the Rule-1 test $\mathfrak{f}^{\mathrm{re}}\gtrless 1/2$ in~\eqref{eq:idx_rules} is the most significant bit of the fractional part. \par

In the interference term~\eqref{eq:idx_interf}, each part of $\tilde x_j$ is written in binary as $\sum_m b_m 2^{m}$, where $b_m\in\{0,1\}$ is the $m$-th bit. The product then becomes $\rho_{i,j}\,\tilde x_j = \sum_m b_m (\rho_{i,j} \ll m)$, a sum of $\lceil\log_2(M{+}1)\rceil = 3$, $4$, and $5$ shifted copies of $\rho_{i,j}$ for $64$-, $256$-, and $1024$-QAM. The constant term of~\eqref{eq:idx_interf} is formed the same way, shifting each $\rho_{i,j}$ by the fixed bit pattern of $M(1+\mathrm{j})$ instead of by $\tilde x_j$. \par

\section{Hardware Architecture}
\label{sec:arch}
Most commercial MIMO receivers use linear ZF or MMSE(-SIC) detection rather than tree search such as $K$-Best. The main reason is the post-quantization datapath of tree-search detectors, i.e., the PED update, candidate expansion and selection after we get the reference node. It is more complex to pipeline than the nearest-point quantization of ZF and MMSE. Box-SCP and Box-SMMECP are feed-forward and fully parallel, so their post-quantization datapath can be simplified while keeping tree-search accuracy. \par

The proposed design is implemented in fixed-point arithmetic for a $4 \times 4$ MIMO detector that supports QAM orders up to $1024$-QAM in the complex domain. \emph{Box-SCP} and \emph{Box-SMMECP} share the same datapath. As in existing $K$-Best ASIC designs~\cite{K-best_JSAC, KBest_VLSI, multigigabit_2014}, the QR decomposition is assumed to be done before detection. The detector produces hard outputs. Its architecture is the skeleton of a soft-output detector, which adds only an LLR calculation module, and soft output is out of the scope of this paper. \par

In this section, all datapath quantities of the proposed design are in the index QAM map $\tilde{\mathcal{A}}$ and carry a tilde, and the PED is the $\ell_1$-norm PED~\eqref{eq:l1_ped}. \par

\subsection{Top-Level Architecture}
\label{sec:arch_top}
\begin{figure*}[t]
    \centering
    \begin{subfigure}{\textwidth}
        \centering
        \includegraphics[width=0.6\textwidth]{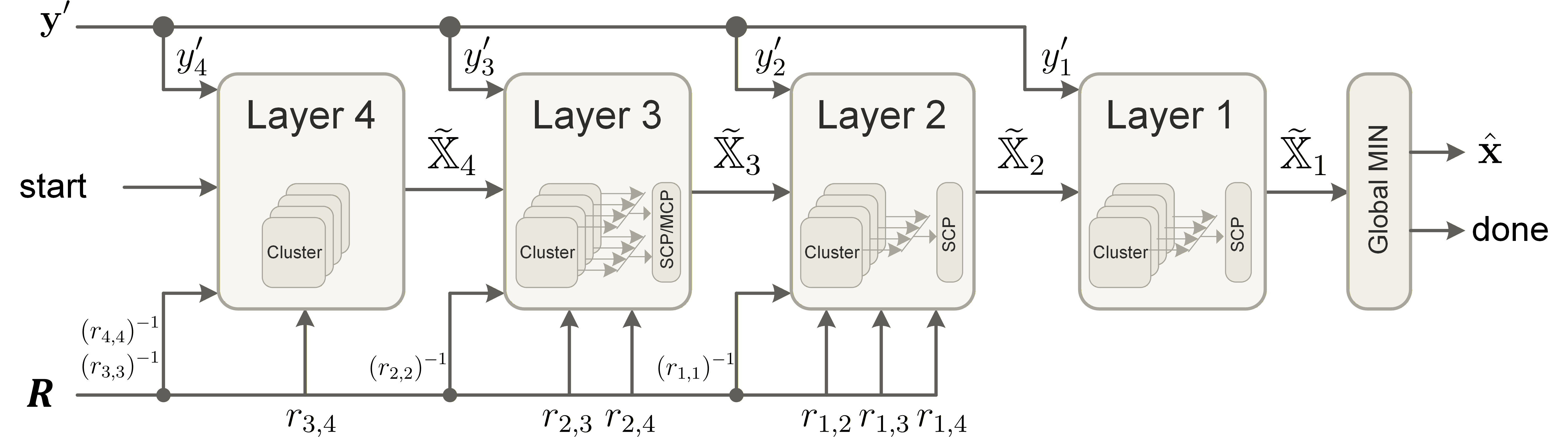}
        \caption{Top-level detector architecture}
        \label{fig:detector_struct}
    \end{subfigure}


    \begin{subfigure}{0.56\textwidth}
        \centering
        \includegraphics[width=0.95\linewidth]{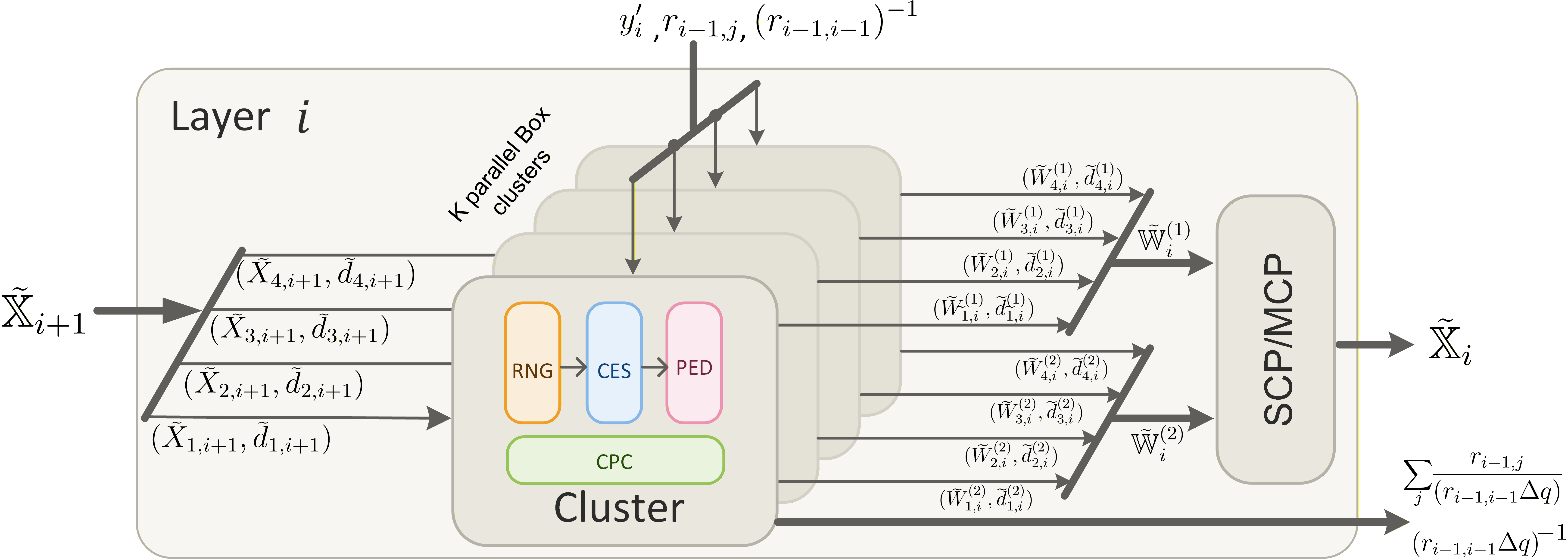}
        \caption{Detection layer level DFG}
        \label{fig:det_layer_dfg}
    \end{subfigure}%
    \begin{subfigure}{0.4\textwidth}
        \centering
        \includegraphics[width=0.95\linewidth]{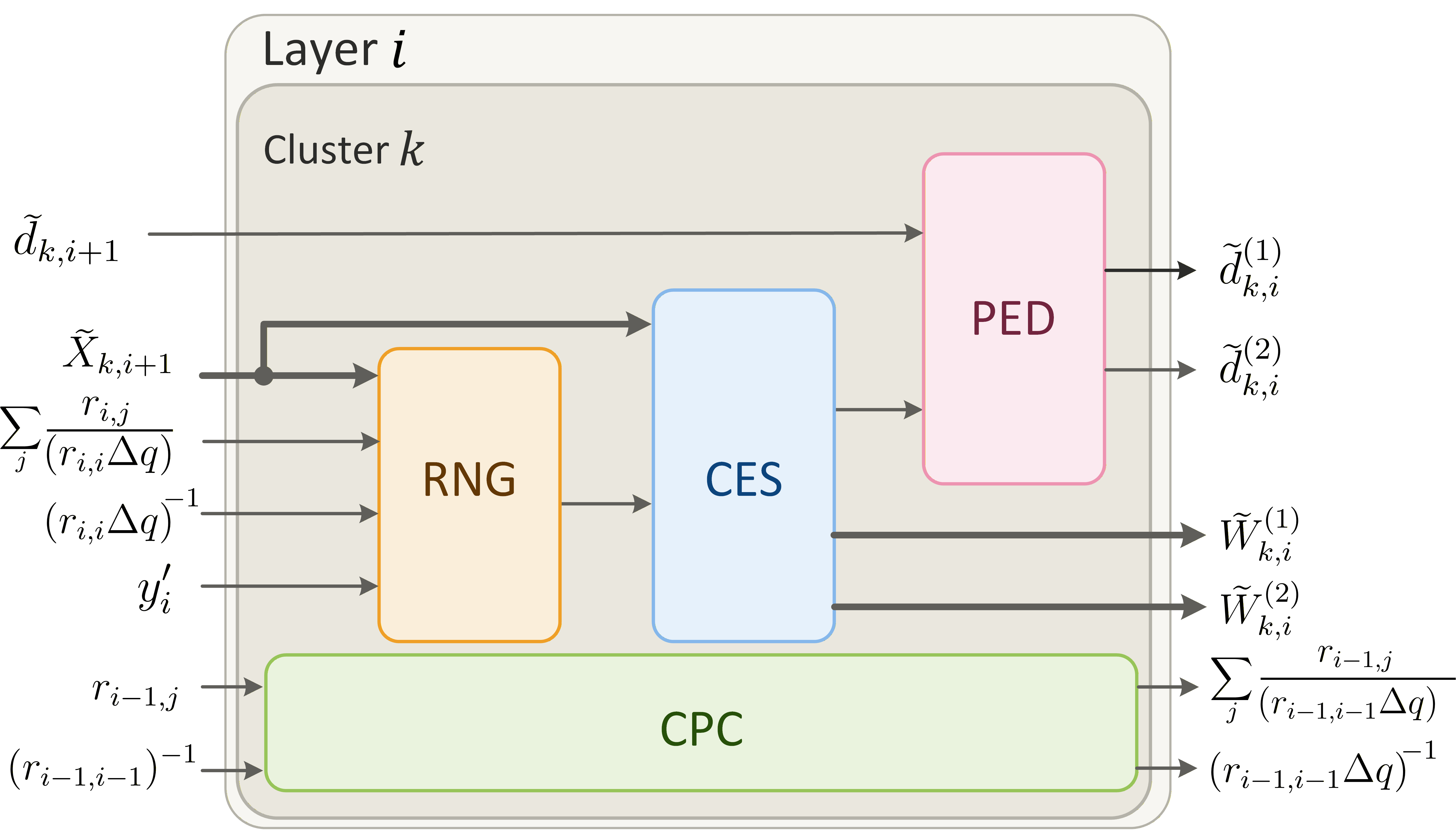}
        \caption{Box cluster level DFG}
        \label{fig:det_cluster_dfg}
    \end{subfigure}
    \caption{Block diagram of the proposed $4 \times 4$ MIMO Box decoder architecture.}
    \label{fig:det_layer_cluster_struct}
\end{figure*}

The detector takes as inputs the upper-triangular matrix $\mathbf{R}$, the QR-rotated received vector $\mathbf{y}'$, the reciprocals $r_{i,i}^{-1}$ of the diagonal terms, and a detection-start flag. It outputs the detected index vector $\tilde{\mathbf{x}}$, which maps to the symbol vector $\hat{\mathbf{x}}$ by~\eqref{eq:idx_sym}, together with a detection-end flag. The reciprocals $r_{i,i}^{-1}$ are forwarded directly from the QR step, where they are already computed to normalize the non-diagonal elements, which avoids division in hardware. \par

Fig.~\ref{fig:detector_struct} shows the detector, which consists of four pipelined detection layers. Detection runs from Layer~4 to Layer~1, top-down through the search tree. The elements of $\mathbf{R}$ and the received symbols $y'_i$ are distributed to the respective layers. In each layer, the augmented path set $\tilde{\mathbb{X}}_i$ is generated and forwarded to the next layer for further expansion and pruning. At Layer~4, candidate symbols are expanded using the Box operation~\eqref{eq:box_def} without pruning. The detector supports two operating modes, Box-SCP and Box-SMMECP. In Box-SCP, every pruning layer (Layers~3, 2, and~1) runs SCP. In Box-SMMECP, Layer~3 becomes the MMECP layer and runs the SCP/MMECP block, while Layers~2 and~1 run SCP, as shown in Fig.~\ref{fig:detector_struct}. After the last layer, a global minimum selection unit identifies the path with the smallest accumulated PED and outputs $\tilde{\mathbf{x}}$. Because SCP and MMECP are feed-forward, the detector is built as a single feed-forward pipeline with no feedback loop. The four detection layers are cascaded, and fine-grained pipeline registers are inserted between the sub-blocks within each layer. A pipeline register is placed at the output of every real multiply--accumulate (MAC). Sub-blocks that contain no MAC but still have a long combinational path, namely the shift--add interference network, the candidate-distance and quantization logic, and the comparator trees of the MMECP layer and the global minimum selection unit, are partitioned by one or two additional registers so that no register-to-register stage exceeds a single MAC delay. The clock period is therefore set by a single MAC delay, and the detector accepts one received vector and outputs one detected vector per cycle. \par

\subsection{Detection Layer and Box Cluster}
\label{sec:arch_layer}
Fig.~\ref{fig:det_layer_dfg} shows a generic detection layer. Each layer $i$ contains $K$ parallel Box clusters operating on the incoming augmented path set $\tilde{\mathbb{X}}_{i+1}$, one Box cluster per parent path $\tilde{X}_{k,i+1}$ with its accumulated PED $\tilde{d}_{k,i+1}$. Each Box cluster $k$ produces the winner paths $\tilde{W}^{(1)}_{k,i}$ and $\tilde{W}^{(2)}_{k,i}$ with their PEDs $\tilde{d}^{(1)}_{k,i}$ and $\tilde{d}^{(2)}_{k,i}$, which are collected into the augmented rank groups $\tilde{\mathbb{W}}^{(1)}_{i}$ and $\tilde{\mathbb{W}}^{(2)}_{i}$. In an SCP layer, $\tilde{\mathbb{W}}^{(1)}_{i}$ directly becomes the new augmented path set $\tilde{\mathbb{X}}_{i}$, so the SCP/MMECP block is only wiring and $\tilde{\mathbb{W}}^{(2)}_{i}$ is not used. In the MMECP layer, the SCP/MMECP block realizes the max-min exchange of Algorithm~\ref{algo: MMECP} with two binary comparator trees and a final PED comparator. Since the datapath is fully feed-forward and pipelined, no operator is reused, so each binary comparator tree is built from $K-1$ comparators, giving $2K-1$ comparators plus the multiplexing logic in total. \par

Fig.~\ref{fig:det_cluster_dfg} shows the data-flow graph (DFG) inside one Box cluster, which contains four blocks, the Coefficient Pre-Computation (CPC), the Reference Node Generation (RNG), the Candidate Expansion and Selection (CES), and the PED Calculation. At layer $i$, the CPC block takes the off-diagonal entry $r_{i-1,j}$ and the reciprocal $(r_{i-1,i-1})^{-1}$ and pre-computes the interference coefficients $\rho_{i-1,j}=r_{i-1,j}/(r_{i-1,i-1}\Delta q)$ and $1/(r_{i-1,i-1}\Delta q)$ used by the next layer $i-1$, so that they are ready before that layer starts (Section~\ref{sec:arch_rng}). The RNG block takes the received symbol $y'_i$, the parent path $\tilde{X}_{k,i+1}$, and the coefficients $\rho_{i,j}$ and $1/(r_{i,i}\Delta q)$ pre-computed by the CPC block of layer $i+1$, and generates the reference node $\tilde{a}_{k,i}$ in the index QAM map $\tilde{\mathcal{A}}$. The CES block expands the Box cluster around $\tilde{a}_{k,i}$, builds Rule~1 and Rule~2 from the distances between the candidates and the reference node, and appends the selected candidates to $\tilde{X}_{k,i+1}$ to form the winner paths $\tilde{W}^{(1)}_{k,i}$ and $\tilde{W}^{(2)}_{k,i}$. The PED block accumulates the winner increments $r_{i,i}\|\tilde\delta^{(b)}\|_1$ of~\eqref{eq:l1_ped} onto the parent PED $\tilde{d}_{k,i+1}$ and outputs $\tilde{d}^{(1)}_{k,i}$ and $\tilde{d}^{(2)}_{k,i}$. \par

\subsection{Reference Node Generation}
\label{sec:arch_rng}
\begin{figure}[t]
    \centering
    \includegraphics[width=0.95\columnwidth]{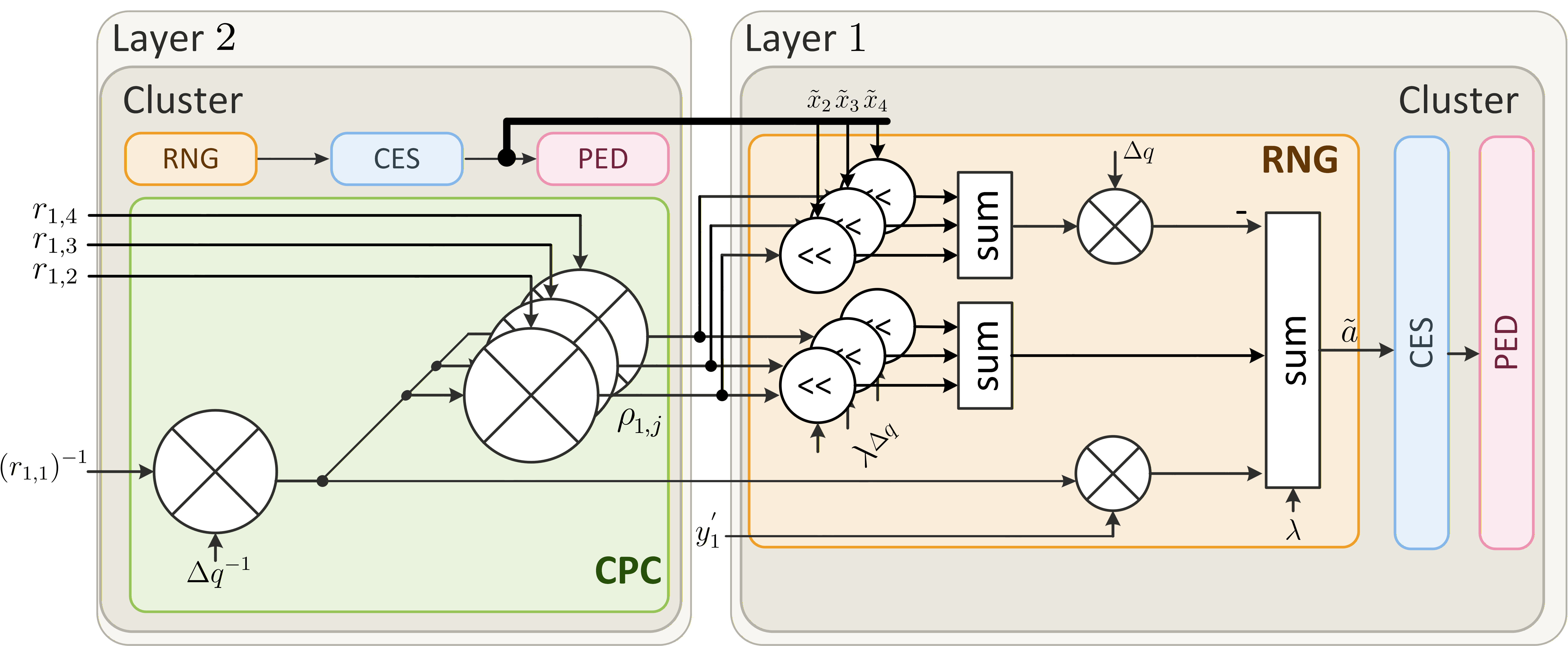}
    \caption{Operator-level DFG of the RNG block.\protect\footnotemark}
    \label{fig:opt_pre_q_dfg}
\end{figure}
\footnotetext{Real and complex operators are not distinguished. The figure shows only the data flow.}
A direct implementation of~\eqref{eq:box_ai_alg} computes the interference term from the previously detected symbols $\hat x_j$ and the off-diagonal elements $r_{i,j}$ of $\mathbf{R}$, subtracts it from $y'_i$, and multiplies the result by $r_{i,i}^{-1}$ to obtain $\hat a$ in $\mathcal{A}$, which is then mapped to $\tilde a$ (Section~\ref{sec:idx_map}). Without pipelining, the critical path of this direct form is about $2T_{\text{CMAC}}$, where $T_{\text{CMAC}}$ is the latency of one complex multiply--accumulate operation. \par

Since $\rho_{i,j}$ and $1/(r_{i,i}\Delta q)$ in~\eqref{eq:idx_a} can be pre-computed (Section~\ref{sec:idx_rng}), their computation is moved out of the per-layer critical path into the CPC block, which runs one layer ahead in parallel with the other blocks. The term $y'_i/(r_{i,i}\Delta q)$ then needs a single multiplication, in parallel with the interference sum, so only MAC operations remain in the per-layer datapath (Fig.~\ref{fig:opt_pre_q_dfg}). The unpipelined per-layer critical path thus drops to about $1\,T_{\text{CMAC}}$. \par

The interference multipliers are replaced by shift--add networks that realize~\eqref{eq:idx_interf} (Fig.~\ref{fig:opt_pre_q_dfg}), which removes them from the per-layer path (Section~\ref{sec:idx_hw}). \par

\subsection{Candidate Expansion and Selection}
\label{sec:arch_ces}
\begin{figure}[!t]
\centering
\includegraphics[width=0.95\columnwidth]{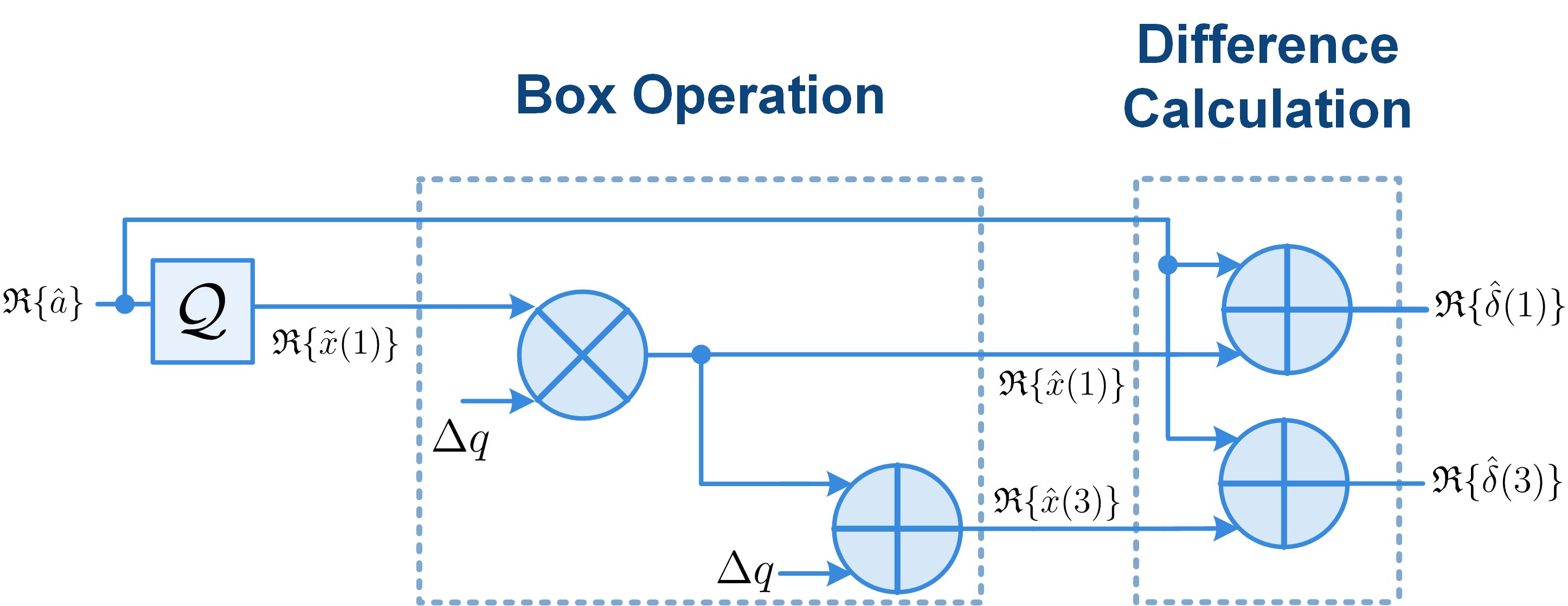}
\caption{Operator-level DFG of the post-quantization processing in the QAM grid $\mathcal{A}$ for the real component.}
\label{fig:ces_removed}
\end{figure}

\begin{figure}[!t]
\centering
\includegraphics[width=\columnwidth]{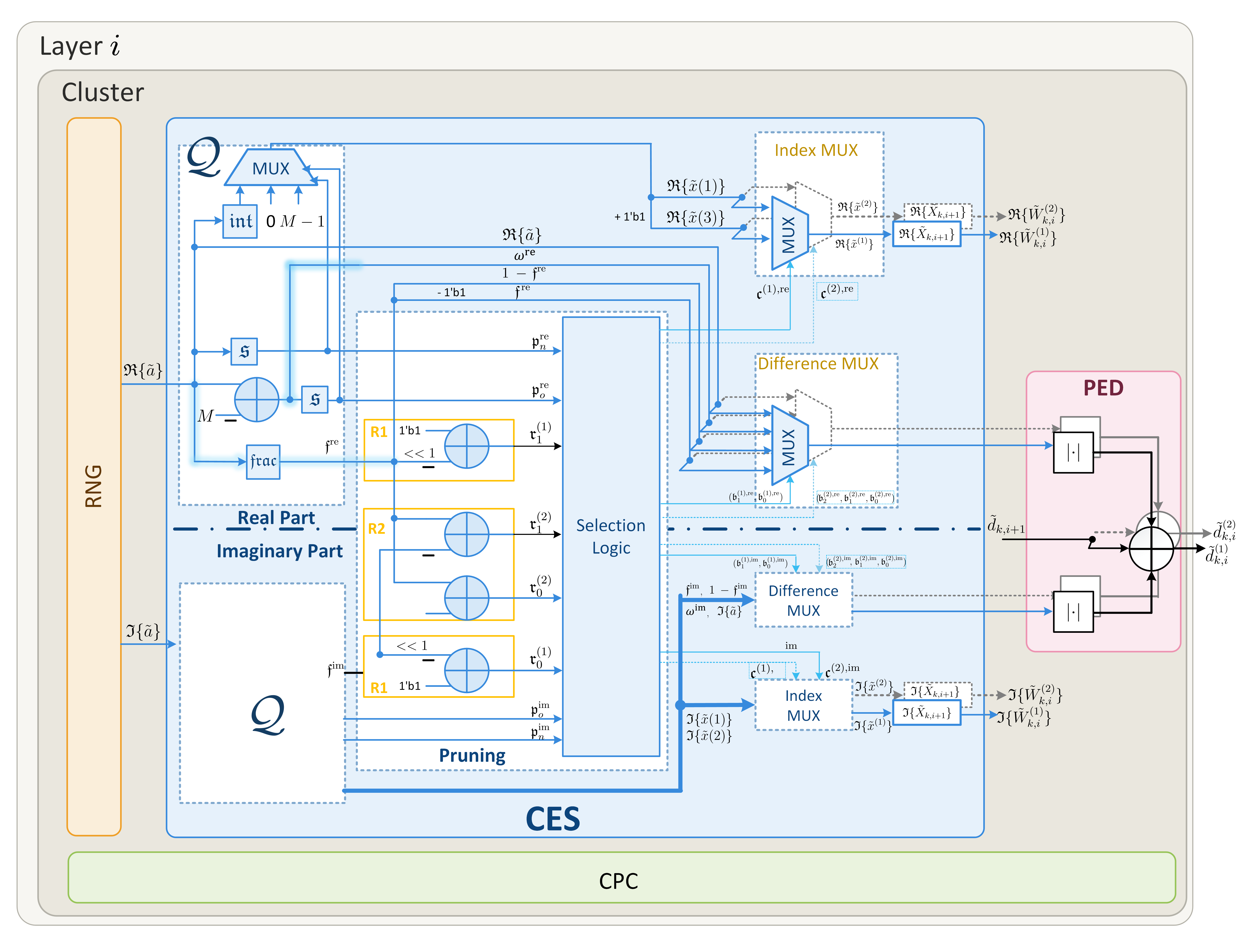}
\caption{Operator-level DFG of the post-quantization processing in the index QAM map $\tilde{\mathcal{A}}$ for the real component.}
\label{fig:post_q_opt_dfg}
\label{fig:post_q_dfg_opt}
\end{figure}
Fig.~\ref{fig:ces_removed} shows the operator-level DFG of the processing after the quantization ($\mathcal{Q}$) in the QAM grid $\mathcal{A}$ for the real component. The reference node $\hat a$ is first mapped to $\tilde a$ and quantized by the $\mathcal{Q}$ block to the starting candidate index $\tilde x(1)$ per the integer assignment of Section~\ref{sec:idx_map}. \par

In a straightforward implementation of~\eqref{eqn:box_operation}, the $\mathcal{Q}$ block output $\mathfrak{R}\{\tilde x(1)\}$ is forwarded to the Box Operation block, which demaps it by multiplying by $\Delta q$ to form $\mathfrak{R}\{\hat x(1)\}$ and adds $\Delta q$ to form $\mathfrak{R}\{\hat x(3)\}$ in $\mathcal A$. As noted in Section~\ref{sec:idx_map}, the offset by $M$ cancels in the difference. The Difference Calculation block then subtracts both coordinates from $\mathfrak{R}\{\hat a\}$ to obtain the distance components $\mathfrak{R}\{\hat\delta(1)\}$ and $\mathfrak{R}\{\hat\delta(3)\}$ for Rule~1, Rule~2, and the PED. \par

Owing to the geometric regularity of the Box, the four candidate coordinates' components $\mathfrak{R}\{\hat x(1)\}$, $\mathfrak{R}\{\hat x(3)\}$, $\mathfrak{I}\{\hat x(1)\}$, $\mathfrak{I}\{\hat x(2)\}$ and their distance components fully describe the Box cluster. The Selection Logic of Section~\ref{sec:arch_sel} picks among them. Without pipeline registers inside the block, the critical path of this straightforward implementation after the $\mathcal{Q}$ block and before the PED block consists of the multiplier and adder of the Box Operation block, the subtractor of the Difference Calculation block, the pruning operations, and the Selection Logic. \par

Fig.~\ref{fig:post_q_opt_dfg} shows the operator-level DFG of the post-quantization processing in the index QAM map $\tilde{\mathcal{A}}$ for the real component. The imaginary component follows the same structure and shares the Rule~2 computation and the Selection Logic. The $\mathcal{Q}$ block realizes the integer assignment of Table~\ref{tab:int_assign} as a three-input multiplexer that selects between the integer part $\mathfrak{i}^{\mathrm{re}}$ of $\mathfrak{R}\{\tilde a\}$, the lower boundary $0$, and the upper boundary $M-1$, controlled by the position indicator $(\mathfrak{p}_n^{\mathrm{re}}, \mathfrak{p}_o^{\mathrm{re}})$ of the region of $\mathfrak{R}\{\tilde a\}$, which are the sign bits of $\mathfrak{R}\{\tilde a\}$ and of the overflow comparator output $\mathfrak{R}\{\tilde a\}-M$ (Section~\ref{sec:arch_sel}). The integer-part case applies $\mathfrak{int}(\cdot)$, which keeps the integer bits and drops the fractional bits of the fixed-point value. \par

With the closed forms of Table~\ref{tab:dis_comp}, the Difference Calculation block reduces to bit selection. In the real dimension, only the distance components $\mathfrak{R}\{\tilde\delta^{(1)}\}$\footnote{As in Section~\ref{sec:surv_analysis}, the superscript $^{(b)}$, $b=1,2$, marks the first and second winners, distinct from the candidate index $(l)$.} and $\mathfrak{R}\{\tilde\delta^{(2)}\}$ of the first and second winners $\tilde x^{(1)}$ and $\tilde x^{(2)}$ in each Box cluster are selected from four quantities already available in the $\mathcal{Q}$ block, namely the fractional bits $\mathfrak{f}^{\mathrm{re}}$ of $\mathfrak{R}\{\tilde a\}$, its complement $1-\mathfrak{f}^{\mathrm{re}}$, the overflow-comparator output $\omega^{\mathrm{re}}=\mathfrak{R}\{\tilde a\}-M$, and the negation $-\mathfrak{R}\{\tilde a\}$, which is the magnitude in the \emph{neg} region, as shown in Fig.~\ref{fig:post_q_opt_dfg}. The selection is driven by the position indicator of $\mathfrak{R}\{\tilde a\}$ and the pruning decision, detailed in Section~\ref{sec:arch_sel}. \par

When implemented in $\tilde{\mathcal{A}}$, the Box Operation and the Difference Calculation blocks are removed. The starting candidate index $\tilde x(1)$ is forwarded directly to the Index MUX, and the distance components are forwarded directly to the Difference MUX (Fig.~\ref{fig:post_q_dfg_opt}). This helps both timing and area. For timing, the critical path after the $\mathcal{Q}$ block and before the PED block reduces to the pruning operations and the Selection Logic. For area, the Box Operation multiplier and adder and the two Difference Calculation subtractors of Fig.~\ref{fig:ces_removed} are replaced by a single increment that forms $\mathfrak{R}\{\tilde x(3)\}$. The imaginary component is processed in the same way, where the increment forms $\mathfrak{I}\{\tilde x(2)\}$.

\subsection{Pruning Operations and Selection Logic}
\label{sec:arch_sel}
\begin{table*}[!t]
    \centering
    \caption{First-winner Selection Logic encodings.}
    \label{tab:m1_enc_tb}
    \begin{subtable}[t]{0.31\textwidth}
    \centering
    \begin{tabular}{@{}ccc@{}}
    \toprule
    $\mathfrak{p}_n$ & $\mathfrak{p}_o$ & Position\\
    \midrule
    1 & 1 & \emph{neg} \\
    0 & 1 & \emph{mid} \\
    0 & 0 & \emph{ovf} \\
    \bottomrule
    \end{tabular}
    \caption{Position-indicator encoding.}
    \label{tab:m1_pos_enc}
    \end{subtable}
    \hfill
    \begin{subtable}[t]{0.31\textwidth}
    \centering
    \begin{tabular}{@{}ccc@{}}
    \toprule
    $\mathfrak{c}^{(1),\mathrm{re}}$ & $\mathfrak{c}^{(1),\mathrm{im}}$ & Selected $\tilde x(l)$ \\
    \midrule
    0 & 0 & $\tilde x(1)$ \\
    0 & 1 & $\tilde x(2)$ \\
    1 & 0 & $\tilde x(3)$ \\
    1 & 1 & $\tilde x(4)$ \\
    \bottomrule
    \end{tabular}
    \caption{Index MUX encoding.}
    \label{tab:m1_idx_enc}
    \end{subtable}
    \hfill
    \begin{subtable}[t]{0.34\textwidth}
    \centering
    \begin{tabular}{@{}cccc@{}}
    \toprule
    $\mathfrak{b}_1$ & $\mathfrak{b}_0$ & $\mathfrak{R}\{\tilde\delta^{(1)}\}$& $\mathfrak{I}\{\tilde\delta^{(1)}\}$\\
    \midrule
    0 & 0 & $\mathfrak{R}\{\tilde a\}$           & $\mathfrak{I}\{\tilde a\}$           \\
    0 & 1 & $\mathfrak{f}^{\mathrm{re}}$         & $\mathfrak{f}^{\mathrm{im}}$         \\
    1 & 0 & $1-\mathfrak{f}^{\mathrm{re}}$       & $1-\mathfrak{f}^{\mathrm{im}}$       \\
    1 & 1 & $\omega^{\mathrm{re}}$                    & $\omega^{\mathrm{im}}$                    \\
    \bottomrule
    \end{tabular}\\[2pt]
    \caption{Difference MUX encoding.}
    \label{tab:m1_dist_enc}
    \end{subtable}
\end{table*}

The pruning operations implement Rule~1 and Rule~2 on the $\Delta q$-normalized displacement $\tilde\delta(1)$, following~\eqref{eq:idx_rules}. The signs of the four inequalities give the rule bits $(\mathfrak{r}^{(1)}_1,\mathfrak{r}^{(1)}_0)$ for Rule~1 and $(\mathfrak{r}^{(2)}_1,\mathfrak{r}^{(2)}_0)$ for Rule~2. \par

The Selection Logic (SL) after the pruning operations determines the surviving candidates of each Box cluster and their distance components. The SL takes as input the four position-indicator bits of the reference node $\tilde a$, namely the real part $(\mathfrak{p}_n^{\mathrm{re}}, \mathfrak{p}_o^{\mathrm{re}})$ and imaginary part $(\mathfrak{p}_n^{\mathrm{im}}, \mathfrak{p}_o^{\mathrm{im}})$, together with the rule bits. The negative flags $\mathfrak{p}_n$ and the overflow flags $\mathfrak{p}_o$ are the sign bits defined in Section~\ref{sec:arch_ces}, so the position indicator reuses existing logic and needs no dedicated comparator. Box-SCP uses only the two Rule-1 bits  $(\mathfrak{r}^{(1)}_1,\mathfrak{r}^{(1)}_0)$ for the first-winner selection (six input bits in total), whereas Box-SMMECP additionally uses the two Rule-2 bits $(\mathfrak{r}^{(2)}_1,\mathfrak{r}^{(2)}_0)$ for the second-winner selection (eight input bits in total). \par

From these inputs, the SL generates the control signals for the index selection and the distance selection MUXs. For the first winner, the control signals $(\mathfrak{c}^{(1),\mathrm{re}}, \mathfrak{c}^{(1),\mathrm{im}})$ select the surviving candidate index in the real and imaginary dimensions, respectively, and drive the Index MUXs of Fig.~\ref{fig:post_q_dfg_opt} with the encoding of Table~\ref{tab:m1_idx_enc}. In parallel, the control signals $(\mathfrak{b}^{(1),\mathrm{re}}_1, \mathfrak{b}^{(1),\mathrm{re}}_0)$ and $(\mathfrak{b}^{(1),\mathrm{im}}_1, \mathfrak{b}^{(1),\mathrm{im}}_0)$ select the corresponding distance components through the Difference MUXs with the encoding of Table~\ref{tab:m1_dist_enc}. Box-SCP uses only these first-winner signals. Box-SMMECP additionally produces the analogous second-winner controls $(\mathfrak{c}^{(2),\mathrm{re}}, \mathfrak{c}^{(2),\mathrm{im}})$ for the Index MUX and the three-bit selects $(\mathfrak{b}^{(2),\mathrm{re}}_2, \mathfrak{b}^{(2),\mathrm{re}}_1, \mathfrak{b}^{(2),\mathrm{re}}_0)$ and $(\mathfrak{b}^{(2),\mathrm{im}}_2, \mathfrak{b}^{(2),\mathrm{im}}_1, \mathfrak{b}^{(2),\mathrm{im}}_0)$ for the Difference MUXs. \par

In each dimension, the position indicator encodes the \emph{neg}, \emph{mid}, and \emph{ovf} regions as $(\mathfrak{p}_n,\mathfrak{p}_o)=11$, $01$, and $00$, respectively (Table~\ref{tab:m1_pos_enc}). The six input bits of the first-winner selection give $2^6=64$ raw input combinations. Most of them are either impossible or carry no information. The pattern $(\mathfrak{p}_n,\mathfrak{p}_o)=(1,0)$ never occurs in either dimension, since $\mathfrak{R}\{\tilde a\}$ cannot be negative and overflowing at the same time, which leaves three valid regions (\emph{neg}, \emph{mid}, \emph{ovf}) per dimension and hence nine position combinations. Within a dimension, the corresponding Rule-1 bit matters only when that dimension lies in the \emph{mid} region, where it selects between the two candidates straddling the reference node. In \emph{neg} and \emph{ovf} the boundary clip already fixes the winner and the Rule-1 bit is a don't-care. The unreachable input patterns and the don't-care Rule-1 bits are treated as don't-cares in the logic minimization, which is carried out with the Logic Friday tool and gives a purely combinational implementation. After minimization, the output control signals are expressed in Boolean form as
\begin{equation}
\begin{aligned}
(\mathfrak{c}^{(1),\mathrm{re}}, \mathfrak{c}^{(1),\mathrm{im}}) &= \left(\overline{\mathfrak{p}_o^{\mathrm{re}}} + \overline{\mathfrak{p}_n^{\mathrm{re}}}\,\mathfrak{r}^{(1)}_1,\; \overline{\mathfrak{p}_o^{\mathrm{im}}} + \overline{\mathfrak{p}_n^{\mathrm{im}}}\,\mathfrak{r}^{(1)}_0 \right), \\
(\mathfrak{b}^{(1),\mathrm{re}}_1, \mathfrak{b}^{(1),\mathrm{re}}_0) &= \left(\overline{\mathfrak{p}_o^{\mathrm{re}}} + \overline{\mathfrak{p}_n^{\mathrm{re}}}\,\mathfrak{r}^{(1)}_1,\; \overline{\mathfrak{p}_o^{\mathrm{re}}} + \overline{\mathfrak{p}_n^{\mathrm{re}}}\,\overline{\mathfrak{r}^{(1)}_1} \right), \\
(\mathfrak{b}^{(1),\mathrm{im}}_1, \mathfrak{b}^{(1),\mathrm{im}}_0) &= \left(\overline{\mathfrak{p}_o^{\mathrm{im}}} + \overline{\mathfrak{p}_n^{\mathrm{im}}}\,\mathfrak{r}^{(1)}_0,\; \overline{\mathfrak{p}_o^{\mathrm{im}}} + \overline{\mathfrak{p}_n^{\mathrm{im}}}\,\overline{\mathfrak{r}^{(1)}_0} \right),
\end{aligned}
\end{equation}
where $\overline{\,\cdot\,}$, $+$, and juxtaposition denote logical NOT, OR, and AND, respectively. Each control signal is a two-level logic function of the input bits. \par

The second-winner paths, used by the SCP/MMECP block, are shown as dashed connections in Fig.~\ref{fig:post_q_dfg_opt}. \par

\subsection{Summary of Architectural Gains}
\label{sec:arch_sum}
The optimizations in $\tilde{\mathcal{A}}$ shorten the RNG critical path to $\sim 1T_{\text{CMAC}}$ through coefficient pre-computation, replace the interference multiplier with shift--add networks, remove the CES Box Operation and Difference Calculation blocks (leaving only the pruning operations and the Selection Logic after the $\mathcal{Q}$ block and before the PED block), and make the Selection Logic purely combinational. \par

These gains bring the post-quantization datapath close to the complexity of MMSE detection, as targeted at the start of Section~\ref{sec:arch}. In an MMSE(-SIC) receiver, the per-symbol work after interference cancellation is a single nearest-point quantization. On top of this quantization, a baseline tree-search detection layer adds a multi-stage sorter network ($K$-Best) or an iterative selection loop (DKB~\cite{DKB_v1}, ICP), whose compare-and-swap stages dominate the per-layer latency. The proposed layer instead adds only one increment for the candidate index, the pruning operations of~\eqref{eq:idx_rules}, and one addition and one multiply--accumulate for the $\ell_1$ PED. \par

\section{Performance Evaluation}
\label{sec:hw_results}
This section evaluates the proposed architecture in two steps. Section~\ref{sec:ces_cmp} compares one detection layer of the different approaches to show the saving of the proposed feed-forward MMECP over the iterations of ICP and the benefit of the new $\tilde{\mathcal{A}}$ hardware datapath. Section~\ref{sec:hw_comparison} compares the full detector with state-of-the-art detectors. All designs of this work are synthesized with Cadence Genus in TSMC $16$\,nm CMOS at a $0.8$\,V supply and a $1$\,GHz clock. The clock is set by the pipeline configuration of this design, which places a register after every real MAC, so the critical path is one real MAC. The $1$\,GHz clock is the synthesis target and also the maximum frequency $f_{\max}$ of this single-MAC critical path in the $16$\,nm technology. The compared detectors are implemented in technology nodes from $65$ to $130$\,nm, so their figures are normalized to $16$\,nm. Power is estimated in Cadence Joules from the switching activity of the synthesized netlist, which is simulated at the gate level in Questa with back-annotated delays at the $1$\,GHz clock. The workload is vectors from a $4\times4$ MIMO system with random bits, an i.i.d.\ Rayleigh channel and an SNR of $20$\,dB, generated with a fixed seed and quantized to the word lengths of Section~\ref{sec:cosim}. The power is averaged over a steady-state window at the maximum throughput of each design. \par

\subsection{Index-Domain Datapath and Pruning Schemes}
\label{sec:ces_cmp}
\label{sec:arch_cmp}
One detection layer with $K=4$ survivor paths is synthesized for each approach at $64$-QAM with the word lengths of Section~\ref{sec:cosim}. Table~\ref{tab:ces_cmp} lists the area and power of the candidate expansion and of the candidate selection, together with the latency and the throughput of the layer. The PED calculation is not listed, since its cost scales with the number of evaluated candidates (Table~\ref{tab:prune_cost}). Following~\cite{multigigabit_2014}, the throughput is
\begin{equation}
\Phi = \frac{N \cdot \log_2 |\mathcal{A}| \cdot f_{\mathrm{clk}}}{N_{\mathrm{clk}}} \quad \mathrm{b/s},
\label{eq:throughput}
\end{equation}
where $N$ is the MIMO size, $|\mathcal{A}|$ the constellation size, $f_{\mathrm{clk}}$ the clock frequency, and $N_{\mathrm{clk}}$ the number of clock cycles per detected vector, i.e., the cycles per layer input of Table~\ref{tab:prune_cost}. \par

For the Box variants, the candidate expansion is the quantization by the $\mathcal{Q}$ block, followed in $\mathcal{A}$ by the Box Operation block (Fig.~\ref{fig:ces_removed}) and in $\tilde{\mathcal{A}}$ by direct wiring (Fig.~\ref{fig:post_q_dfg_opt}). The datapath in $\tilde{\mathcal{A}}$ never maps the candidate back to $\mathcal{A}$, so each $\tilde\delta(l)$ is read directly from the bits of $\tilde a$. The two SCP columns of Table~\ref{tab:ces_cmp} therefore show the saving of the $\tilde{\mathcal{A}}$ datapath. In $\tilde{\mathcal{A}}$, the candidate expansion takes $10$\,\textmu m$^2$ instead of $51$\,\textmu m$^2$, and the candidate selection $494$\,\textmu m$^2$ instead of $602$\,\textmu m$^2$, so both blocks together need $23\%$ less area and $44\%$ less power at the same latency of $3$ cycles. For DKB, the candidate expansion is the Schnorr-Euchner enumeration, which takes $65$\,\textmu m$^2$. \par

For SCP and MMECP, the candidate selection consists of the pruning operations, the Selection Logic, and the MUXs, plus the max-min exchange for MMECP, which is fully feed-forward. ICP and DKB~\cite{DKB_v1} select the survivors in $K$ dependent rounds (Section~\ref{sec:complexity}). For both, an area-oriented version folded into one loop and a throughput-oriented version unfolded into $K$ round units are implemented for comparison. In the $1$\,GHz implementation, each selection round takes three clock cycles, so the latency of ICP is $13$ cycles in both forms, and the folded versions accept a new layer input every $3K=12$ cycles and reach $2$\,Gb/s by~\eqref{eq:throughput}. The candidate selection of MMECP takes $1324$\,\textmu m$^2$ and $8.1$\,mW, i.e., $73\%$ less area and $75\%$ less power than the unfolded ICP at the same throughput of $24$\,Gb/s, with a latency of $5$ instead of $13$ cycles. The folded ICP needs $15\%$ less area than MMECP but reaches only one twelfth of its throughput. This confirms the trade-off of Section~\ref{sec:complexity}, where the iterative selection of ICP costs either area or throughput, while MMECP reaches the full throughput with a small area. DKB needs more area than ICP in both forms, $1608$\,\textmu m$^2$ folded and $8203$\,\textmu m$^2$ unfolded, with a latency of $26$ cycles. \par

\begin{table}[!t]
\centering
\caption{Area, power, latency, and throughput of candidate expansion and candidate selection for one layer ($64$-QAM, $16$\,nm, $K{=}4$, $1$\,GHz).}
\label{tab:ces_cmp}
\scriptsize
\setlength{\tabcolsep}{2.5pt}
\begin{tabular}{@{}lccccccc@{}}
\toprule
 & \multirow{2}{*}{SCP,\,$\mathcal{A}$} & \multirow{2}{*}{SCP,\,$\tilde{\mathcal{A}}$} & \multirow{2}{*}{MMECP} & \multicolumn{2}{c}{ICP} & \multicolumn{2}{c}{DKB~\cite{DKB_v1}}\\
\cmidrule(lr){5-6}\cmidrule(lr){7-8}
 & & & & folded & unfolded & folded & unfolded\\
\midrule
\multicolumn{8}{@{}l}{\textit{Candidate expansion}}\\
Area (\textmu m$^2$) & 51 & 10 & 10 & 10 & 10 & 65 & 65\\
Power (mW)           & 0.3 & ${<}0.1$ & ${<}0.1$ & ${<}0.1$ & ${<}0.1$ & 0.3 & 0.3\\
\midrule
\multicolumn{8}{@{}l}{\textit{Candidate selection}}\\
Area (\textmu m$^2$) & 602 & 494 & 1324 & 1124 & 4988 & 1608 & 8203\\
Power (mW)           & 5.2 & 3.1 & 8.1 & 1.0 & 32.4 & 1.9 & 57.4\\
\midrule
\multicolumn{8}{@{}l}{\textit{Total}}\\
Area (\textmu m$^2$) & 653 & 504 & 1334 & 1134 & 4998 & 1673 & 8268\\
Latency (cycles)     & 3 & 3 & 5 & 13 & 13 & 26 & 26\\
Throughput (Gb/s)    & 24 & 24 & 24 & 2 & 24 & 2 & 24\\
\bottomrule
\end{tabular}\\[2pt]
{\scriptsize MMECP and ICP are implemented in $\tilde{\mathcal{A}}$, DKB in $\mathcal{A}$.}
\end{table}

\subsection{Comparison with State-of-the-Art}
\label{sec:hw_comparison}
\label{sec:synth}
The gate count is reported in NAND2-equivalent gates (kGE). Tables~\ref{tab:cmp_64qam} and~\ref{tab:cmp_highqam} list the results for both Box-SCP and Box-SMMECP across the three modulation orders. One received vector is detected per cycle, so $N_{\mathrm{clk}}=1$ in~\eqref{eq:throughput}, which gives $24$, $32$, and $40$\,Gb/s for $64$-, $256$-, and $1024$-QAM, respectively. The pipeline latency is $19$ cycles, i.e., $19$\,ns at $1$\,GHz.


The proposed detector is compared with published $K$-Best ASICs, grouped by modulation order. Table~\ref{tab:cmp_64qam} lists $4\times4$ $64$-QAM designs, and Table~\ref{tab:cmp_highqam} lists $4\times4$ high-order ($256$/$1024$-QAM) designs. Only designs that report the gate count, the throughput, and the power are listed, so that every normalized metric can be computed. Since these designs use different technology nodes, supply voltages, and $K$ values, normalized metrics are given with the raw figures. Following~\cite{multigigabit_2014}, the normalized hardware efficiency (NHE, in Mbps/kGE) and normalized power efficiency (NPE, in Gb/J) are
\begin{equation}
\mathrm{NHE}=\frac{\Phi\cdot(t/t_0)}{G},\qquad
\mathrm{NPE}=\frac{\Phi\cdot(t/t_0)}{P\,(1/V_{dd})^2},
\label{eq:nhe_npe}
\end{equation}
where $\Phi$ is the throughput, $G$ the gate count, $P$ the power, $V_{dd}$ the supply voltage, and $t$ the process node. The throughput is linearly projected from $t$ to a common node $t_0=16$\,nm. The projection favors the references, so Tables~\ref{tab:cmp_64qam} and~\ref{tab:cmp_highqam} also list the raw ratio $\Phi/G$.

All proposed designs operate at $K=B=4$. At this point, Box-SMMECP is within about $0.2$\,dB of $K$-Best with the same $K$, and Box-SCP within about $0.8$\,dB (Section~\ref{sec:ber}). The references use $K=8$ to $16$ and reach a lower BER, so the tables compare designs at different operating points. The figure of merit (FOM) of~\cite{radius_adaptive_2010} is defined for this comparison. It normalizes the throughput by the gate count per surviving node $G/K$ to remove the influence of $K$. The original FOM uses the raw throughput, which is inconsistent with the technology projection applied to the NHE and NPE. We therefore project the throughput to the common node $t_0$ in the FOM as well, which gives
\begin{equation}
\mathrm{FOM}=\frac{1}{C}\cdot\frac{\Phi\cdot(t/t_0)}{G/K},
\label{eq:fom}
\end{equation}
with $C=1$ for the QR preprocessing used by all listed designs. The proposed-design columns report the synthesized Box-SCP and Box-SMMECP datapaths at $16$\,nm. Box-SCP is the lowest-cost mode, and Box-SMMECP is the higher-accuracy mode for about $6\%$ more gates. NPE is listed only where the source reports both supply voltage and power.

\begin{table*}[!t]
\centering
\caption{Comparison of $4\times4$ Hard-Output MIMO Detectors}
\label{tab:cmp_64qam}
\scriptsize
\setlength{\tabcolsep}{2pt}
\begin{tabular}{l c c c c c c c c}
\hline
& \cite{radius_adaptive_2010} & \cite{k_best_sort_free} & \cite{DKB_v1} & \cite{multigigabit_2014} & \cite{KBest_VLSI} & \cite{sort_free_2024} & This work & This work \\
\hline
Modulation       & 64-QAM & 64-QAM & 64-QAM & 64-QAM & 64-QAM & 64-QAM & 64-QAM & 64-QAM \\
Algorithm        & DKB & DKB & DKB & DKB & DKB & $K$-Best & Box-SCP & Box-SMMECP \\
Signal domain    & Real & Real & Complex & Real & Real & Real & Real & Real \\
Output           & Hard & Hard & Hard & Hard & Hard & Hard & Hard & Hard \\
Process (nm)     & 65 & 65 & 130 & 90 & 90 & 90 & 16 & 16 \\
$K$              & 8 & 10 & 10 & 8 & 10 & 16 & 4 & 4 \\
Max. freq. (MHz) & --- & 158 & 282 & 170 & 102 & 200 & 1000 & 1000 \\
Gate count (kGE) & 210 & 1760 & 114 & 232 & 251 & 853.3 & 103.9 & 110.5 \\
Core area (mm$^2$) & --- & --- & 2.31 & 0.92 & 0.77 & --- & 0.022 & 0.023 \\
Throughput (Mb/s) & 252.6 & 463 & 675 & 4080 & 489 & 4800 & 24000 & 24000 \\
Vectors/s (M) & 10.5 & 19.3 & 28.1 & 170 & 20.4 & 200 & 1000 & 1000 \\
Power (mW)       & 12.8 & 165 & 135 & 234 & 17 & 224 & 103.9 & 110.2 \\
Implementation   & Synthesis & Synthesis & Silicon & Silicon & Silicon & Synthesis & Synthesis & Synthesis \\
$\Phi/G$ (Mbps/kGE) & 1.2 & 0.3 & 5.9 & 17.6 & 1.9 & 5.6 & 231.0 & 217.2 \\
\hline
NHE @16\,nm (Mbps/kGE) & 4.9 & 1.1 & 48.1 & 98.9 & 11.0 & 31.6 & 231.0 & 217.2 \\
NPE @16\,nm (Gb/J) & --- & 11.4 & 68.7 & 165.8 & 161.8 & 145.8 & 147.8 & 139.4 \\
FOM @16\,nm (Mbps$\cdot K$/kGE) & 39.1 & 10.7 & 481.1 & 791.4 & 109.6 & 506.3 & 924.0 & 868.8 \\
\hline
\end{tabular}

\vspace{2pt}
\begin{minipage}{\textwidth}
\scriptsize
Dashes (---) denote values not reported in the source.
\end{minipage}
\end{table*}

\begin{table*}[!t]
\centering
\caption{Comparison of $4\times4$ High-Order QAM MIMO Detectors}
\label{tab:cmp_highqam}
\footnotesize
\setlength{\tabcolsep}{4pt}
\begin{tabular}{l c c c c c c}
\hline
& \cite{sort_free_2024} & This work & This work & \cite{sort_free_2024} & This work & This work \\
\hline
Modulation        & 256-QAM & 256-QAM & 256-QAM & 1024-QAM & 1024-QAM & 1024-QAM \\
Algorithm         & $K$-Best & Box-SCP & Box-SMMECP & $K$-Best & Box-SCP & Box-SMMECP \\
Output            & Hard & Hard & Hard & Hard & Hard & Hard \\
Process (nm)      & 90 & 16 & 16 & 90 & 16 & 16 \\
$K$               & 16 & 4 & 4 & 16 & 4 & 4 \\
Max. freq. (MHz)  & 200 & 1000 & 1000 & 200 & 1000 & 1000 \\
Gate count (kGE)  & 1842 & 134.6 & 142.9 & 2697 & 169.7 & 179.7 \\
Throughput (Mb/s) & 6400 & 32000 & 32000 & 8000 & 40000 & 40000 \\
Vectors/s (M) & 200 & 1000 & 1000 & 200 & 1000 & 1000 \\
Power (mW)        & 483 & 152.4 & 170.6 & 707 & 213.3 & 228.3 \\
Implementation    & Synthesis & Synthesis & Synthesis & Synthesis & Synthesis & Synthesis \\
$\Phi/G$ (Mbps/kGE) & 3.5 & 237.7 & 223.9 & 3.0 & 235.7 & 222.6 \\
\hline
NHE @16\,nm (Mbps/kGE) & 19.5 & 237.7 & 223.9 & 16.7 & 235.7 & 222.6 \\
NPE @16\,nm (Gb/J) & 90.2 & 134.4 & 120.0 & 77.0 & 120.0 & 112.1 \\
FOM @16\,nm (Mbps$\cdot K$/kGE) & 312.7 & 951.0 & 895.7 & 267.0 & 942.8 & 890.4 \\
\hline
\end{tabular}
\end{table*}

Table~\ref{tab:cmp_64qam} compares the $64$-QAM designs. Box-SCP reaches the highest throughput of $24$\,Gb/s with the smallest gate count of $103.9$\,kGE, and thus the highest normalized hardware efficiency (NHE) of $231.0$\,Mbps/kGE, about $2.3\times$ that of the next-best detector~\cite{multigigabit_2014}. Box-SMMECP follows with $217.2$\,Mbps/kGE. The gain comes either from removing the per-layer sorter or selection loop or from operating at $K=4$ rather than $K=8$ to $16$. To remove the influence of $K$, the FOM normalizes the gate count per survivor, and Box-SCP reaches $924$\,Mbps$\cdot K$/kGE, about $1.2\times$ that of~\cite{multigigabit_2014}. The NPE of Box-SCP is $147.8$\,Gb/J, about $11\%$ below that of~\cite{multigigabit_2014}.

Table~\ref{tab:cmp_highqam} compares the $256$- and $1024$-QAM designs with the sort-free detector of~\cite{sort_free_2024}, the only listed design at these orders. Box-SCP reaches $32$ and $40$\,Gb/s with $134.6$ and $169.7$\,kGE, which gives an NHE of $237.7$ and $235.7$\,Mbps/kGE, about $12\times$ and $14\times$ that of~\cite{sort_free_2024}, and a FOM of $951$ and $943$\,Mbps$\cdot K$/kGE, about $3.0\times$ and $3.5\times$. The NPE of the proposed designs is $1.3$ to $1.6\times$ that of~\cite{sort_free_2024}.

The gate count grows only mildly with the modulation order, rising about $63\%$ from $64$- to $1024$-QAM ($103.9$ to $169.7$\,kGE for Box-SCP, $110.5$ to $179.7$\,kGE for Box-SMMECP), since the datapath stays in $\tilde{\mathcal{A}}$ and only the word lengths widen. Over the same range, the gate count of the sort-free detector of~\cite{sort_free_2024} grows by $216\%$, from $853.3$ to $2697$\,kGE, because the comparisons of its candidate search and its pipeline registers grow with the constellation. Box-SMMECP adds only the single-layer MMECP block, for gate-count and power overheads of $6.4$/$6.2$/$5.9\%$ and $6.0$/$11.9$/$7.0\%$ at $64$/$256$/$1024$-QAM.

To relate the raw throughput to an application, the tables also list detected vectors per second, $\Phi/(N\log_2|\mathcal A|)$. A $5$G NR carrier of $100$\,MHz with $30$\,kHz subcarrier spacing and four layers needs about $92$\,M vectors/s, and a Wi-Fi~7 link of $320$\,MHz with four streams needs about $290$\,M vectors/s. Among the compared designs, only this work exceeds this vector rate in one instance, for the Wi-Fi~7 modulations up to $1024$-QAM. The surplus of a $1$\,GHz instance can be traded for a lower clock and supply voltage, or one instance can be shared across carriers.

\section{Conclusion}
This paper presented an iteration-free pruning scheme and a VLSI architecture for sort-free MIMO Box Decoding in high-order QAM systems. The MMECP scheme replaces the sequential selection loop of ICP by a single max-min exchange derived from survival statistics. It keeps $K$ paths per layer with $2K-1$ comparators and no feedback, and matches $K$-Best performance. The index-domain datapath keeps the entire post-quantization processing in the index QAM map $\tilde{\mathcal{A}}$, so quantization, box expansion and distance extraction become bit-field selection. This removes the operations needed for candidate expansion and selection on symbol values, and the selection logic becomes one fixed combinational function for every modulation order. In one $64$-QAM detection layer, the $\tilde{\mathcal{A}}$ datapath reduces the area of candidate expansion and selection by $23\%$ and the power by $44\%$, and MMECP needs $73\%$ less area and $75\%$ less power than unfolded ICP at the same throughput. Synthesized for a $4\times4$ detector in $16$\,nm CMOS, the design detects one vector per cycle at $1$\,GHz, sustains $24$, $32$ and $40$\,Gb/s at $64$-, $256$- and $1024$-QAM with $63\%$ gate-count growth, and reaches $2.3\times$, $12\times$ and $14\times$ the normalized hardware efficiency of the best compared $4\times4$ hard-output detector at $64$-, $256$- and $1024$-QAM. Soft output and larger arrays are the next steps.

\bibliographystyle{IEEEtran}
\bibliography{references}

\end{document}